\documentclass[12pt]{iopart}
\usepackage{graphicx}
\usepackage{rotating}

\begin{document}
\title{Exact Christoffel-Darboux Expansions: A New, Multidimensional, Algebraic, Eigenenergy Bounding Method}
\author{Carlos R. Handy}
\address{Department of Physics, Texas Southern University, Houston, Texas 77004}
\ead{carlos.handy@tsu.edu}

\begin{abstract}
Although the Christoffel-Darboux representation (CDR) plays an important role within the theory of orthogonal polynomials, and many important bosonic and fermionic  multidimensional Schrodinger equation systems can be transformed into a moment equation representation (MER), the union of the two into an effective, algebraic, eigenenergy bounding method has been overlooked. This particular fusion of the two representations, suitable for bounding bosonic or fermionic systems, defines the Orthonormal Polynomial Projection Quantization - Bounding Method (OPPQ-BM), as developed here. We use it to analyze several one dimensional and two dimensional systems, including the quadratic Zeeman effect for strong-superstrong magnetic fields. For this problem, we match or surpass the excellent, but intricate, results of Kravchenko et al (1996 {\em Phys. Rev. A} {\bf 54} 287) for a broad range of magnetic fields, without the need
for any truncations or approximations.  
\end{abstract}

\newpage
\pagestyle{plain}

\section{Introduction}

It is well recognized that the development of effective bounding methods for generating tight (converging) lower and upper bounds to the discrete state energies is an important problem. This is because many systems,  particularly those exhibiting strong coupling interactions, involve significant multiscale dynamics mandating the use of singular perturbation methods [1]. These specialized methods, including the adapation of conventional methods (i.e. large order nonorthogonal basis expansions, large order perturbation resummation analysis, asymptotic analysis, etc.), may yield varying and/or inaccurate results, motivating the need for tight bounds by which to gauge the reliability of competing estimation methods. 

Many important, low dimension, Schrodinger equation eigenenergy problems are transformable into a moment equation representation (MER). That is, their power moments will satisfy a linear recursion relation with the energy as a parameter. Within this context, the Eigenvalue Moment Method (EMM) [2-4] was developed by Handy et al, and proved to be an effective eigenenergy bounding method, generating geometrically converging bounds, for strongly coupled systems such as the quadratic Zeeman effect for strong - superstrong magnetic fields [3-7]. However, since the method depends on positive, or nonnegative, configuration space representations for the discrete states [2-4,8-10], and exploits the well known positivity theorems arising from the {\it Moment Problem} in mathematics [11], it is presently limited only to solving for the multidimensional bosonic ground state. Presently, EMM cannot be extended to multidimensional excited bosonic states, or fermionic systems. Additionally, the EMM algorithms are based on the use of convex optimization methods such as semidefinite programming (SDP)  [12,13] and linear programming [3,4,14], which are not  traditional mathematical techniques within physics. 

Despite the limitations of EMM, it has long been the objective to develop other MER based bounding methods capable of addressing multidimensional  excited bosonic states and fermionic systems. In this we have been succesful, the focus of this work. This advance is achieved through the realization that by embedding the Christoffel-Darboux (basis) representation (CDR) [15] within a MER formulation, the expansion coefficients can be generated in closed form, resulting in a bounding theory capable of generating converging bounds for all discrete states. Furthemore, the computational implementation is purely algebraic with no truncations or approximations. We refer to this bounding formulation as the Orthonormal Polynomial Projection Quantization-Bounding Method (OPPQ-BM).

Despite the present limitations of EMM, we expect it to be more efficient  (i.e. the same bounding accuracy using less power moments) than OPPQ-BM since it focuses on the pointwise positivity of the bosonic ground state wavefunction, and not just on the positivity of a particular integral expression. However, the computational implementation of OPPQ-BM is far simpler than EMM, and can be done to (essentially) arbitrary precision utilizing advanced algebraic software, such as Mathematica.

The fusion of CDR and MER is not new. This was the essence of the eigenenergy approximation formalism developed by Handy and Vrinceanu [16,17]. We shall refer to this approach as the Orthonormal Polynomial Projection Quantization - Approximation Method (OPPQ-AM). Despite its effectiveness as an estimation method, it overlooked the fact that its basic structure, when viewed from a different  analytical perspective, leads to a converging, eigenenergy, bounding method. 

In Sec. 2, we provide an overview of the essential structure of the OPPQ-BM formalism. Approximately half of this is a review of the OPPQ-AM formalism, upon which OPPQ-BM is partly based. This is followed by implementation of OPPQ-BM on the quantum harmonic oscillator (Sec. 3), the quartic anharmonic oscillator (Sec. 4), and the two dimensional quadratic Zeeman problem (QZM) in Sec. 5. The effectiveness of OPPQ-BM is vindicated by our ability to match and surpass the eigenenergy estimates  by Kravchenko et al [6] and Schimerczek and Wunner [7], for a broad range of magnetic field strengths, through an algebraic procedure involving no truncations or other approximations (i.e. B-splines, etc.). Additionally, we can generate eigenenergy bounds to the ground and first excited states within the $L_z = 0$, even parity, symmetry class ($0^+$).  Each problem illustrates the general structure of the formalism for one and multidimensional systems. Interspersed within the specific examples, we provide proofs, and other rationale, for important relations. In the Appendix, we provide those important proofs not given in the earlier sections. 
\newpage
\section{Overview of the OPPQ-BM Formalism}

The following overview pertains to the OPPQ-BM formalism, developed within the context of a one space dimension formulation. The basic structure is unchanged for multidimensional systems, unless noted otherwise.

One can skip to Sec. 3 and Sec. 4, for the harmonic oscillator and the quartic anharmonic oscillator, respectively, which develop the underlyng formalism while solving the corresponding problem.  

\subsection{ Abbreviated Overview}

For any one dimensional or multidimensional system, we have:

{\bf Step 1}: Given the Christoffel-Darboux representation  (CDR) for an unknown discrete state wavefunction (Eq.1), and assuming the given quantum system admits a Moment Equation representation (i.e. MER, Eq.(4)), we can generate the CDR expansion coefficients in closed (exact) form (Eq.(5)). 

{\bf Step 2}: From the OPPQ-BM quantization condition in Eq.(12), we can generate a purely energy dependent function, ${\cal L}_I(E)$, whose asymptotic properties in the expansion order ``$I$" (Eq.(14)) lead to both the generation of eigenenergy estimates  (i.e. Eq.(16), different from OPPQ-AM), and converging eigenenergy bounds (Eqs.(21-22)). The expression ${\cal L}_I(E)$ results from a constrained quadratic form minimization (CQFM) ansatz (i.e. Eq.(10)).

{\bf Step 3}: For one dimensional systems, ${\cal L}_I(E) = \lambda_I(E)$, corresponding to the smallest eigenvalue of a certain positive definite matrix (Eq.(8)); whereas for multidimensional problems the CQFM ansatz must consider alternative constraints to those for one space dimension problems.

\subsection{Comprehensive Overview}

\subsubsection{\underline{The Christoffel-Darboux Representation}:}

Let $R(x) > 0$ be a positive, exponentially decaying weight with finite power moments, $w(p) = \int dx \ x^p R(x)$. Let $P_n(x) = \sum_{j=0}^n\Xi_j^{(n)} x^j$ be its orthonormal polynomials satisfying $\langle P_m|R|P_n\rangle = \delta_{m,n}$.  Consider the decomposition  of the discrete state wavefunction, $\Psi$, in terms of the non-orthogonal basis $\{P_n(x) R(x)|n\geq 0\}$:
\begin{eqnarray}
\Psi(x) = \sum_{n=0}^\infty c_n P_n(x) R(x),
\end{eqnarray}
where the projection coefficients are given by
\begin{eqnarray}
c_n & = & \langle P_n|\Psi\rangle, \\ \cr
& = & \sum_{j=0}^n\Xi_j^{(n)} \mu(j),
\end{eqnarray}
involving the power moments, $\mu(p) = \int dx \ x^p\Psi(x)$. The expansion in Eq.(1)  defines the Christoffel-Darboux representation (CDR), although Handy and Vrinceanu [16,17] referred to it as the Orthonormal Polynomial Projection Quantization (OPPQ) representation. 

In their eigenenergy estimation analysis, referred to here as the OPPQ-Approximation Method (OPPQ-AM),  it was argued that the better the weight is modeled after the asymptotic form of the physical states, the faster convergent will be the CDR/OPPQ expansion in Eq.(1). This can involve the use of Freud type weights (i.e. $R(x) = exp(-\gamma\ x^{2q})$, etc.), and consideration of completeness issues. In their work on the sextic anharmonic oscillator [16], the use of the Freud weight $exp({-{1\over 4}x^4})$ produced excellent results. In this work, we will use combinations of the classical orthogonal polynomials; therefore completeness issues will not be of concern. We note that the above basis $\{P_n(x) R(x)\}$ is non-orthogonal; however, it is a linear combination of an orthonormal basis formed from the orthogonal polynomials relative to the weight $R^2$.

\subsubsection{\underline{The Moment Equation Representation (MER)}:}

Assume that the physical system admits a linear recursion relation for the power moments, of the form
\begin{eqnarray}
\mu(p) = \sum_{\ell = 0}^{m_s}M_E(p,\ell) \ \mu_{\ell},
\end{eqnarray}
for $p \geq 0$, where the initialization moments, otherwise referred to as the {\it missing moments},  correspond to $\mu_\ell = \mu(\ell)$, for $0 \leq \ell \leq m_s$.  The coefficients, $M_E(p,\ell)$, are known functions of the energy, $E$, satisfying $M_E(\ell_1,\ell_2) = \delta_{\ell_1,\ell_2}$. Many important physical systems admit such moment equation representations (MER). For one dimensional systems, $m_s = finite$; while for multidimensional systems, $m_s = \infty$, although the missing moments define an infinite hierarchy of nested moment subspaces. That is, given the first $1+m_s$ missing moments, the quantum system is exactly projected within the ${\cal U}_{m_s}$ subspace; and a finite number of dependent moments uniquely generated.

\subsubsection{\underline{Combining CDR and MER: Generating Closed Form Expressions for the ${c_n}$'s}:}
Upon substituing the MER relation into Eq.(3), the CDR projection coefficients will take on the closed (exact) form
\begin{eqnarray}
c_n(E,{\overrightarrow \mu}) = {\overrightarrow \Lambda_E^{(n)}}\cdot {\overrightarrow \mu},
\end{eqnarray}
where ${\overrightarrow \mu} \equiv (\mu_0,\ldots, \mu_{m_s})$, the missing moment vector; and ${\overrightarrow { \Lambda_E^{(n)}}}$ are known energy dependent vectors.

Define the positive partial sums
\begin{eqnarray}
{\cal S}_I(E,{\overrightarrow \mu}) \equiv \sum_{n=0}^I c_n^2(E, {\overrightarrow \mu}).
\end{eqnarray}
It is straightforward to represent these partial sums as the expectation value of an energy dependent, symmetric, positive definite matrix, ${\cal P}_I(E) > 0$, with respect to the missing moment vector:
\begin{eqnarray}
{\cal S}_I(E,{\overrightarrow \mu}) = \langle {\overrightarrow \mu}|{\cal P}_I(E)|{\overrightarrow \mu}\rangle.
\end{eqnarray}
The explicit form for these positive definite matrices are defined for each problem considered in the following sections.
 
Define the smallest eigenvalue for these positive definite matrices:
\begin{eqnarray}
\lambda_I(E) = Smallest\ Eigenvalue \ of \ {\cal P}_I(E).
\end{eqnarray}
For one dimensional systems, all $\{{\cal P}_I(E)|I \geq m_s\}$ will have the same dimension, $1+m_s$; consequently, their smallest eigenvalues
 generate a positive, increasing sequence:
\begin{eqnarray}
 0 < \lambda_I(E) < \lambda_{I+1}(E) <  \ldots < \lambda_{\infty}(E).
\end{eqnarray}
These relations are not generally valid for multidimensional systems because the corresponding positive definite matrices have increasing dimension, as the number of missing moments (i.e. the effective expansion order)  is increased.  

\subsubsection{\underline{Constrained Quadratic Form Minimization (CQFM)}:}

The constrained quadratic form minimization  (CQFM) problem corresponds to:
\begin{eqnarray}
{\cal L}_I(E) \equiv Inf_{\overrightarrow \mu}\{{\cal S}_I(E,{\overrightarrow \mu})|{\cal C}({\overrightarrow \mu}) = 1\}.
\end{eqnarray}
The constraint relation can be linear (to be used in the QZM case) or nonlinear. Clearly, if ${\cal C}({\overrightarrow \mu}) \equiv |{\overrightarrow \mu}|^2$, we recover the eigenvalue functions. In the Appendix we show that these expressions, so long as the constraint holds at the expansion order ``$I$",  will also satisfy the positive, increasing, sequence in Eq.(9):
\begin{eqnarray}
 0 < {\cal L}_I(E) < {\cal L}_{I+1}(E) <  \ldots < {\cal L}_{\infty}(E).
\end{eqnarray}
In the multidimensional case,  the missing moment order $m_s$ becomes the expansion order,  $I\rightarrow m_s$. This CQFM formulation allows us to extend OPPQ-BM to the multidimensional case, for suitable constraints, as discussed in Sec. 5 and in the Appendix. In particular, the linear constraint equivalent to $\mu_0 = 1$, will suffice for the QZM problem with respect to the $0^+$ states.

For one dimensional systems, the more natural constraint is the unit normalization for the missing moment vector; therefore: ${\cal L}_I(E) \equiv \lambda_I(E)$.

For the harmonic oscillator problem discussed in Sec. 3, $m_s = 0$, and the ${\cal P}_I(E)$ positive definite matrix is just a number; therfore ${\cal P}_I(E) \equiv \lambda_I(E)$.  We use the notation $S_I(E) \equiv \lambda_I(E)$ in this case.

For the quartic anharmonic oscillator disccused in Sec. 4, $m_s = 1$, and the underlying positive definite matrix is two dimensional. For this problem, we explicitly use the $\lambda_I(E)$ notation.  

All the properties of the $\lambda_I(E)$ functions, as presented in Sec. 3 (i.e. $\lambda_I(E) \equiv S_I(E)$), or Sec. 4,  apply in the same way to the ${\cal L}_I(E)$ functions as used for multidimensional problems. It is these properties that enable the OPPQ-BM formalism to generate eigenenergy estimates and converging eigenenergy bounds.

\subsubsection{\underline{The OPPQ-AM and OPPQ-BM Quantization Condition}:}

The most important relation within OPPQ-BM is the quantization condition (assuming ${\overrightarrow \mu } \neq {\overrightarrow 0}$):
\begin{eqnarray}
\hspace{-40pt} \langle \Psi|{1\over R}|\Psi \rangle ={\cal S}_\infty(E,{\overrightarrow \mu}) = \cases { finite, \iff E = E_{phys}\ and \ {\overrightarrow \mu} = {\overrightarrow \mu_{phys}} \cr
\infty, \iff E \neq E_{phys}\ or \ {\overrightarrow \mu} \neq {\overrightarrow \mu_{phys}} \cr}.
\end{eqnarray}
This is satisfied only if the weight does not asymptotically go to zero faster than the probability density for the physical states [16,17]; therefore the ``finiteness" condition in Eq.(12) is satisfied, for physical states.  The unbounded asymptotic limit for unphysical states then follows. 

We note that the asymptotic relation in Eq.(12) is also satisfied if the asymptotic form of the weight is that of the physical state 
(i.e. $Lim_{|x| \rightarrow \infty } {{\Psi}\over R} = const$) since the other factor, $\Psi$, will exponentially decay, resulting in a finite integral. If this is the case, then the corresponding integral for unphysical configuration space solutions will become infinite. 

The work by Handy and Vrinceanu [16,17] focuses on Eq.(12), solely for the physical states, concluding that 
\begin{eqnarray}
Lim_{n \rightarrow \infty} c_n(E_{phys},{\overrightarrow \mu}_{phys}) = 0.
\end{eqnarray}
They use this asymptotic condition, at high order, to approximate the physical energies and missing moments. This quantization procedure defines the Orthonormal Polynomial Projection Quantization - Approximation Method (OPPQ-AM), as designated in this work. 

The OPPQ-AM method works well as an estimation method, with faster convergence if the weight is modeled after the asymptotic form of the physical states. However, spurious complex energy roots, with asymptotically vanishing imaginary parts, may contribute. An alternative quantization strategy was required in order to avoid such spurious energies. The key to realizing an alternative quantization strategy (i.e. OPPQ-BM) is to focus on the full extent of Eq.(12) for both physical and unphysical energy and missing moment parameters. 

This capacity of moment representations for discriminating between physical and unphysical solutions is also an important component of EMM. In configuration space, unphysical solutions do not have finite power moments, and therefore are ``filtered" out from the MER relation. However, within the moment representation, there will be  unphysical moment solutions that cannot correspond to any physical state in configuration space. Their unphysical nature is conveyed through the unbounded asymptotic limit given in Eq.(12).
 
\subsubsection{\underline{The OPPQ-BM Ansatz: Energies and Converging Bounds}}

The challenge was to understand how to solve Eq.(12). We can argue that instead of dealing with Eq.(12), we can focus on the simpler problem devoid of any missing moment vectors:

\begin{eqnarray} 
Lim_{I\rightarrow \infty}{\cal L}_I(E) = \cases { finite, \iff E = E_{phys} \cr
\infty, \iff E \neq E_{phys}\cr}.
\end{eqnarray}

We prove this relation, for the case ${\cal L}_I(E) \equiv \lambda_I(E)$, within the discussion on the quartic-anharmonic oscillator, which corresponds to an $m_s = 1$ system. The more general case (i.e. ${\cal L}_I(E) \neq \lambda_I(E)$, due to other types of  constraints, as indicated in Eq.(10)) is presented in the Appendix. The relation in Eq. (14), combined with Eq.(11),  are the key to the generation of converging eigenenergy bounds within the OPPQ-BM formalism. Everything presented below follows from these two relations. 

We repeat the previous point because it is the essence of OPPQ-BM for any system in any dimension. Given Eq.(14), and Eq.(11), in that order, one can generate eigenenergy bounds through OPPQ-BM, for any one dimensional, or multidimensional, problem admitting a MER representation. In Sec. 4, for the quartic anharmonic oscillator, we prove it  in the context of setting ${\cal L}_I(E) \equiv \lambda_I(E)$. In the Appendix, we prove it for the general case, although we apply it to the QZM problem in Sec. 5. The proofs are straightforward.

Clearly, Eq.(14) is telling us that the physical solutions are the local minima of ${\cal L}_\infty(E)$:
\begin{eqnarray}
\partial_E{\cal L}_\infty(E_{phys}) = 0;
\end{eqnarray}
therefore, one should focus on the local minima of the $I$-th order  function, since these should approximate the physical energies to $I$-th order:
\begin{eqnarray}
\partial_E{\cal L}_I(E_I^{(min)}) = 0.
\end{eqnarray}
From Eq.(11), these local minima will in turn  generate another positive, increasing, bounded from above sequence:
\begin{eqnarray}
0 < {\cal L}_I(E_I^{(min)}) < {\cal L}_{I+1}(E_{I+1}^{(min)}) < \ldots < {\cal L}_\infty(E_{phys}) = finite.
\end{eqnarray}
That is, the sequence of local extrema do indeed converge to the physical energies:
\begin{eqnarray}
Lim_{I \rightarrow \infty}E_I^{(min)} = E_{phys}.
\end{eqnarray}

Let ${\cal B}_U$ be any coarse upper bound to the limiting form of the bounded sequence in Eq.(17). If this  sequences converges sufficiently fast, one can determine ${\cal B}_U$. Accordingly, assume that this coarse upper bound has been determined:
\begin{eqnarray}
{\cal B}_U > {\cal L}_\infty(E_{phys}) = finite,\ from\  Eq.(17).
\end{eqnarray}
Solve for the roots:
\begin{eqnarray}
{\cal L}_I(E_I^{(L)}) ={\cal L}_I(E_I^{(U)}) = {\cal B}_U,
\end{eqnarray}
which will always have a solution,  due to Eq.(14). These roots will then define lower and upper bounds to the physical energies:
\begin{eqnarray}
E_I^{(L) }< E_{phys} < E_I^{(U)},
\end{eqnarray}
with
\begin{eqnarray}
Lim_{I\rightarrow \infty}\Big(E_I^{(U)}- E_I^{(L) }\Big) = 0^+.
\end{eqnarray}

In the sections that follow, we implement the above on the quantum harmonic oscillator problem, the quartic anharmonic oscillator, and QZM. Within their presentations, we will offer proofs, or rationales, as needed. In the  Appendix we provide a proof for the general relation in Eq.(14), although it will be similar to that in  Sec. 4, for the case ${\cal L}_I(E) = \lambda_I(E)$. We also provide a proof for the general form of Eq.(11).

\newpage

\section {The Quantum Harmonic Oscillator}

The quantum harmonic oscillator is an $m_s = 0$ problem, when restricted to each symmetry class. We can then set the zeroth order moment to unity, $u_0 = 1$, making the $S_I(E)$ functions in Eq.(6) independent of any missing moments. Thus, effectively, $\lambda_I(E) = {\cal S}_I(E)$; however, for the harmonic oscillator problem we do not, explicitly, use the $\lambda_I(E)$ notation. The OPPQ-BM bounding structure will  develop in a straightforward manner. All the properties exhibited by the $S_I(E)$ functions for the harmonic oscillator problem presented here, will apply to the general $m_s \neq 0$ case, but for their corresponding, purely energy dependent, functions (i.e. $\lambda_I(E)$ and ${\cal L}_I(E)$ as introduced in Sec. 1).

\subsection {The Moment Equation Representation (MER)}

Consider the harmonic oscillator, 
\begin{eqnarray}
-\partial_x^2 \Psi(x) + x^2\Psi(x) = E \Psi(x).
\end{eqnarray}
To transform it into MER form multiply both sides  by $x^p$, and integrate by parts, assuming that $\Psi$ is a discrete state configuration.  We obtain the Hamburger moment equation representation:
\begin{eqnarray}
\mu(p+2) = E \mu(p) +p(p-1) \mu(p-2),
\end{eqnarray}
for $p \geq 0$. This is, effectively, a finite difference equation of order 2, in which the initialization moments $\{\mu(0),\mu(1)\}$ must be specified before all the other moments can be generated, for any  energy parameter value, $E$. We refer to these initialization moments as the {\it missing moments}. For the full  harmonic oscillator (working with the even and odd states simultaneously), the missing moment order is  $m_s = 1$. We also note that the order of the finite difference equation does not change if the kinetic energy term (i.e. $p(p-1)\mu(p-2)$) is removed. This suggests that within a moments representation, or equivalently, a Fourier space representation, singular-perturbation expansions  in configuration space become (more) regular perturbation expansions. 

The MER relation can be rewritten as 
\begin{eqnarray}
\mu(p) = \sum_{\ell = 0}^{m_s = 1} {\tilde M}_E(p,\ell)\mu_\ell,
\end{eqnarray}
for $ p \geq 0$, $\mu_\ell \equiv \mu(\ell)$, and ${\tilde M}_E(\ell_1,\ell_2) = \delta_{\ell_1,\ell_2}$. The energy dependent coefficients ${\tilde M}_E(p,\ell)$ (polynomials in $E$) also satisfy the same MER relation in Eq.(24) with respect to the $p$-index, subject to the indicated initialization conditions.

Restricting our analysis to the symmetric states, for simplicity, the odd order Hamburger moments become zero, $\mu(odd) = 0$. The even order Hamburger moments become Stieltjes moments of a configuration restricted to the nonnegative real axis.  Thus, let $u(p) \equiv \mu(2p) =  \int_0^\infty d\xi \ \xi^p \Phi(\xi)$ where $\Phi(\xi) \equiv {{\Psi(\sqrt{\xi})}\over {\sqrt{\xi}}}$. The new MER relation is 
\begin{eqnarray}
u(p+1) = E\ u(p)+2p(2p-1)\ u(p-1),
\end{eqnarray}
$\ p \geq 0$. This is an $m_s = 0$ problem since only one missing moment is required, $u(0) \equiv u_0$. We then have the  relation (expressed in the standard form)
\begin{eqnarray}
u(p) = \sum_{\ell = 0}^{m_s = 0}M_E(p,\ell) u_{\ell},
\end{eqnarray}
for $p \geq 0$, and $M_E(0,0) = 1$.    The $M_E(p,\ell)$ energy dependent coefficients satisfy Eq.(26) with respect to the $p$-index.

\subsection{The Christoffel-Darboux Representation}

Since the Stieltjes moments, $u(p)$, are the power moments of a function on the nonnegative real axis, $\Phi(\xi)$, $\xi \geq 0$, we can expand the physical solutions in terms of the orthonormal polynomials of an appropriate weight, ${\cal R}(\xi)$.  Given that the asymptotic form for $\Psi(x)$ is governed by $R(x) = exp(-{{x^2}\over 2})$, whose orthogonal polynomials are the  Hermite polynomials $He_n(x)$ satisfying $\langle He_m|R(x)|He_n\rangle = {\cal N}_n \delta_{m,n}$, the transformation $x\rightarrow \xi = x^2$ will involve the weight ${\cal R}(\xi) \equiv {{exp(-\xi/2)}\over\sqrt{\xi}}$. The corresponding orthogonal polynomials would be $He_{2\eta}(\xi^{1\over 2}) $ which become polynomials of degree $\eta$ in the $\xi$ variable. Let us represent their orthonormal counterparts by \begin{eqnarray}
P_\eta(\xi) = \sum_{j=0}^\eta \Xi_j^{(\eta)} \xi^j,
\end{eqnarray}
where
\begin{eqnarray}
\langle P_{\eta_1}|{\cal R}|P_{\eta_2}\rangle = \delta_{\eta_1,\eta_2}.
\end{eqnarray}

The Christoffel-Darboux representation (CDR) expands the physical configuration in terms of the basis $\{P_\eta(\xi){\cal R}(\xi)| \eta \geq 0\}$:

\begin{eqnarray}
\Phi(\xi) = \sum_{\eta = 0}^\infty c_\eta P_\eta(\xi){\cal R}(\xi).
\end{eqnarray}
We emphasize that although the polynomials are orthonormal relative to the chosen weight, the basis functions $P_\eta(\xi){\cal R}(\xi)$ are non-orthogonal relative to each other.

The projection coefficients are generated in closed form from the associated MER relation:
\begin{eqnarray}
c_\eta (E) & = & \langle P_\eta|\Phi\rangle, \ \cr 
& = & \sum_{j=0}^\eta \Xi_j^{(\eta)} u(j), \ \cr
& = & \sum_{j=0}^\eta \Xi_j^{(\eta)} \Big( \sum_{\ell = 0}^{m_s} M_E(j,\ell) \ u_{\ell} \Big), \ \cr
& = & \sum_{\ell = 0}^{m_s = 0} \Lambda_{E,\ell}^{(\eta)} \ u_{\ell},\ \cr
& = & \Lambda_{E,0}^{(\eta)},
\end{eqnarray}
where we have exaggerated the notation in anticipation of the more general case, and taken $u_0 \equiv 1$.
The $\Lambda$-coefficients are given by
\begin{eqnarray}
 \Lambda_{E,\ell }^{(\eta)} = \sum_{j=0}^{\eta}\Xi_j^{(\eta)} M_E(j,\ell ).
\end{eqnarray}

For the chosen weight, the orthonormal polynomials are given through closed form expressions:

\begin{eqnarray}
 P_\eta(\xi) = 
  {{(-{1\over 2})^\eta}
} {{( (2\eta)!)^{1\over 2}}\over{(2\pi)^{1\over 4}}} \sum_{j=0}^\eta{{(-2)^j}\over{(\eta-j)!(2j)!}}\xi^j .
\end{eqnarray}

\subsection{Orthogonal Polynomial Projection Quantization - Approximation Method (OPPQ-AM)}

In their combined CDR-MER/OPPQ-AM eigenenergy estimation formalism, Handy and Vrinceanu [16,17] argued that for the physical solutions, if the chosen weight satisfies the condition that the ratio $Lim_{\xi \rightarrow \infty}{{\Phi_{phys}^2(\xi)}\over {{\cal R}(\xi)}} \rightarrow 0$ (i.e. asymptotically vanishes fast enough so that its integral is finite), then the finiteness of the ensuing positive series 
\begin{eqnarray}
\langle {\Phi}_{phys}|{1\over {\cal R}}|\Phi_{phys}\rangle = \sum_{\eta = 0}^\infty c_\eta^2(E_{phys}) < \infty,
\end{eqnarray}
leads to the quantization condition:
\begin{eqnarray}
Lim_{\eta \rightarrow \infty} c_\eta(E_{phys}) = 0.
\end{eqnarray}
Thus, depending on the missing moment order, $m_s$, for a given expansion order, $N$, one sets to zero the $c_{\eta_\ell}$ coefficients for $\eta_\ell = N-\ell$, and $0 \leq \ell \leq m_s$. Since the harmonic oscillator has $m_s = 0$, then all that is required is to examine the roots of $c_N(E) = 0$, for increasing expansion order, $N$.

Indeed, for the harmonic oscillator problem, the OPPQ-AM ansatz is exact, since one can show that the CDR coefficients assume the form
\begin{eqnarray}
c_\eta(E) = \cases{ {1\over {(2\pi)^{1\over 4}}}   , \ for\ \eta = 0 , \cr
{\cal N}_\eta \Pi_{j=1}^{\eta}(E-(1+4(j-1))), \ for \ \eta \geq 1 . \cr     }
\end{eqnarray}
the roots being the exact even parity state energy values, $E_{2N} = 1+4N$, for $N = 0,1,\ldots$. 

\subsection{The Quantization Condition for  OPPQ-BM} 

From  Eq.(34), we are motivated to consider the partial sums:
\begin{eqnarray}
{\cal S}_I(E) \equiv \sum_{\eta = 0}^I c_\eta^2(E),
\end{eqnarray}
which define a positive, increasing, sequence 
\begin{eqnarray}
0 < {\cal S}_I(E) < {\cal S}_{I+1}(E) < \ldots < {\cal S}_\infty(E),
\end{eqnarray}
and according to Eq. (34) satisfy the OPPQ-BM quantization condition:
\begin{eqnarray}
Lim_{I\rightarrow \infty}{\cal S}_{I}(E) = \cases{ finite, \iff \ E = E_{phys} ;\cr 
\infty, \iff \ E \neq E_{phys} \cr}.  
\end{eqnarray}

Indeed, Eq.(39) holds for all one dimensional and multidimensional problems, provided the proper energy functions are used to define ${\cal S}_I(E)$ (i.e. ${\cal L}_I(E)$ from Eq.(10)).

\begin{figure}
\includegraphics{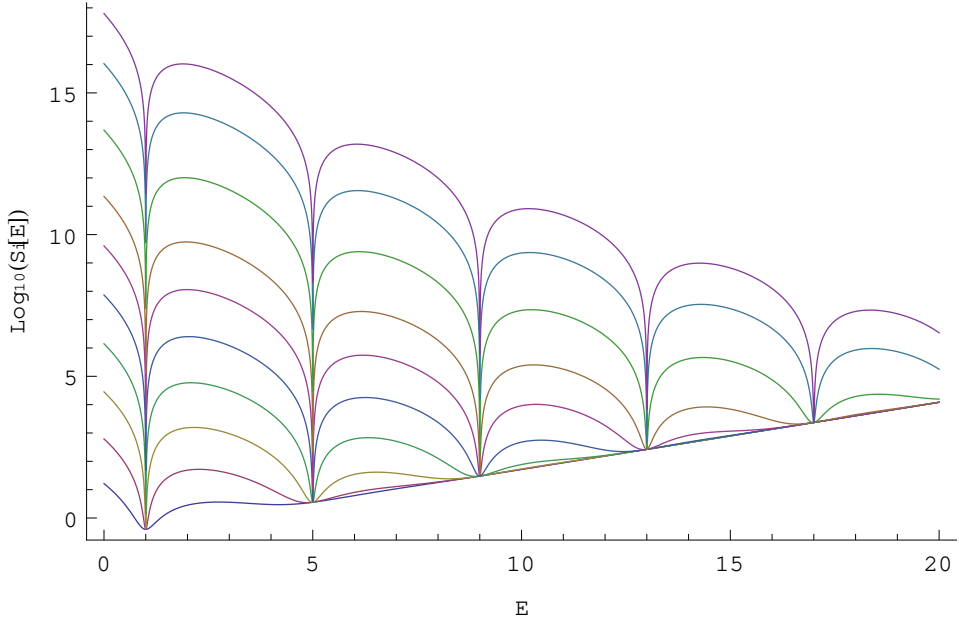}
\centering{\caption{ $Log_{10}(S_I(E))$ for symmetric states of the harmonic oscillator; $I = 5,8,11,\ldots, 32$.}}
\end{figure}

In Figs. 1-3 we illustrate the validity of Eq.(39) over different energy intervals. 

From Eq.(39) an eigenenergy bounding ansatz emerges. Clearly, the essence of Eq.(39) is that the local minima of ${\cal S}_\infty(E)$, correspond to the physical energies;
\begin{eqnarray}
\partial_E\Big( {\cal S}_\infty(E_{phys})\Big) = 0.
\end{eqnarray}
This suggest that to finite order, the corresponding local minima approximate the physical energies. 

Define
\begin{eqnarray}
\partial_E\Big( {\cal S}_I(E_I^{(min)})\Big) = 0.
\end{eqnarray}
Then from Eq.(38) it follows that these local minima satisfy

\begin{eqnarray}
{\cal S}_I(E_I^{(min)}) <  {\cal S}_I (E_{I+1}^{(min)})  < {\cal S}_{I+1}(E_{I+1}^{(min)}) ,
\end{eqnarray}
assuming that $E_{I+1}^{(min)}$ lies within the minima extremal neighborhood of $E_I^{(min)}$, which it will. This means that the local minima define another positive, increasing, bounded from above, sequence:

\begin{eqnarray}
0 < {\cal S}_I(E_I^{(min)}) < {\cal S}_{I+1}(E_{I+1}^{(min)}) < \ldots < {\cal S}_{\infty}(E_{phys}) < \infty.
\end{eqnarray}

\subsection{OPPQ-BM: A High Accuracy Energy  Estimation Method}

As previoulsy noted, one of the problems of the OPPQ-AM approach is that spurious complex energies may appear; although the imaginary parts generally vanish in the infinite order expansion limit. Within OPPQ-BM, the energy approximants (i.e. the local minima) are always real. Furthermore, from Eq.(43), the limit of the local minima is the physical energy:

\begin{eqnarray}
Lim_{I\rightarrow \infty} E_I^{(min)} = E_{phys}.
\end{eqnarray}

\subsection{OPPQ-BM: The Eigenenergy Bounding Process}

Let ${\cal B}_U$ be any coarse upper bound to the local minima sequence:
\begin{eqnarray}
{\cal B}_U > \{ {\cal S}_I(E_I^{(min)})| I \geq 0\}.
\end{eqnarray}
It is implicitly assumed that the sequence elements all correspond to a particular physical energy.
Usually, the increasing positive sequence, ${\cal S}_I(E_I^{(min)})$, converges fast enough, allowing for a quick estimate of the coarse upper bound, ${\cal B}_U$.

From Eq.(39) it follows that at expansion order $I$, there will be roots to the equations
\begin{eqnarray}
{\cal S}_I(E_I^{(L)}) = {\cal S}(E_I^{(U)}) = {\cal B}_U.
\end{eqnarray}
The interval $[E_I^{(L)},E_I^{(U)}]$ must contain the physical energy; thereby generating bounds:
\begin{eqnarray}
E_I^{(L)} < E_{phys} < E_I^{(U)}.
\end{eqnarray}
Furthermore, in the infinite expansion limit, these lower and upper bounds must converge to each other:

\begin{eqnarray}
Lim_{I\rightarrow \infty}\Big( E_I^{(U)} - E_I^{(L)} \Big) = 0.
\end{eqnarray}

\subsection{OPPQ-BM Numerical Results for the Harmonic Oscillator}

In Fig. 1, we plot $Log_{10} \Big( S_I(E)\Big)$  over the interval $0 \leq E \leq 20$.  The nesting of the $S_I(E)$ curves is readily apparent, consistent with Eq.(38) and Eq.(39). Although these functions are nested within each other, their local minima do not necessarily coincide (i.e. as clearly shown in Fig.3 for the second excited state). 

In Fig. 2  we show the progression of localized concavity around the ground state energy ($E_{gr} = 1$)  for the lower order partial sums, $\{S_I(E)| 3 \leq I \leq 12\}$.  Indeed, the $S_I(1)$ sequence is $S_0(1) = S_1(1) = \ldots = {1\over{\sqrt{2\pi}}} = .398942$; thereby concluding, within our OPPQ-BM formulation, that the ground state energy is precisely 1. 

\begin{figure}
\includegraphics{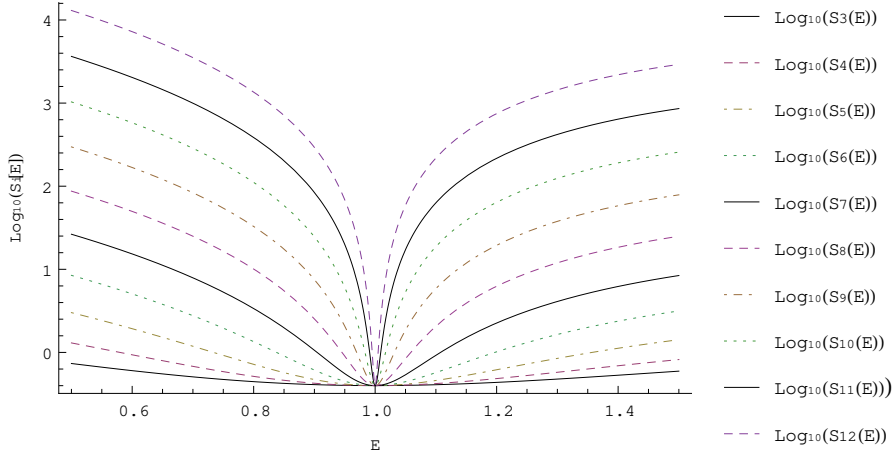}
\centering{\caption{Nesting of the partial sums $Log_{10}(S_I(E))$ centered around the ground state (i.e. $E_{gr} = 1$) for the harmonic oscillator, where $I = 3,4,5,\ldots, 12$. Note that all curves share the same, fixed, minimum.}}
\end{figure}

Things are more interesting for the second excited state, as given in Fig. 3 and  Table I 
. We determine the local minima  $\partial_E S_I(E_I^{(min)}) = 0$, and generate the sequence $\{S_I(E_I^{(min)})\}$ whose convergence  defines ${S}_\infty(E_2) = 3.5904805$. Inspection of the sequence in the third column leads us to conclude that convergence is already setting in at the fourth decimal place. A coarse upper bound ${\cal B}_U = 3.6 > {\cal{S}}_\infty(E_2) $ then allows us to generate converging bounds to the excited state by taking $S_I(E_I^{(L)}) = S_I(E_I^{(U)}) = {\cal  B}_U$, for $I \rightarrow \infty$. Note that the coarseness of the upper bound estimate for ${\cal B}_U$ does not determine the tightness of the eigenenergy bounds (which depend only on the expansion order $I$).

\begin{figure}
\includegraphics{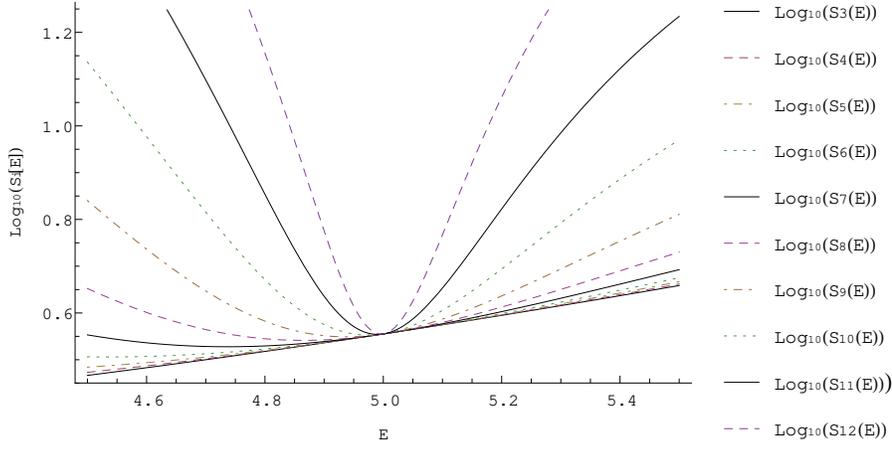}
\centering{\caption{Nesting of the partial sums $Log_{10}(S_I(E))$  centered around the $2^{nd}$ excited state, $E_2 = 5$, for the harmonic oscillator, where $I = 3,4,5,\ldots, 12$. Note that their respective minima, in the energy variable,  monotonically increase to $E= 5$; and they all have the same derivative at that point.}}
\end{figure}

\begin{table}
\caption{
OPPQ-BM  for $E_2$:  $V(x) = x^2$, $R=e^{-{{x^2}\over2}}$ 
}
\centerline{
\begin{tabular}{lcccc}
\hline
$I$ & $\partial_ES_I(E_I^{(min)})=0$    &  $S_I( E_I^{(min)})$  & $E_I^{(L)}$     &  $E_I^{(U)}$     \\
\hline
\hline
 6 & 4.53222&  3.20587&  4.07088&  5.00593\\
 7 & 4.73661& 3.37132&  4.48590&  5.00591 \\
 8 & 4.86462&  3.47875&  4.73214&  5.00585\\
 9 & 4.93802&  3.54002&  4.87312& 5.00572 \\
 10 & 4.97454&  3.56996&  4.94437& 5.00541 \\
 11 & 4.99037&  3.58276&  4.97612& 5.00479 \\
 12 & 4.99656&  3.58773&  4.98933&  5.00384 \\
 13 & 4.99882&  3.58954&  4.99489&5.00276   \\
 14 & 4.99961&  3.59017&   4.99741& 5.00181  \\
\hline
20 &   4.9999996 &3.5904802 &4.99993 & 5.00007\\
\hline
 $\infty$ & 5&  $3.5904805< {\cal B}_U = 3.6$ &  5& 5 \\
 \hline
\end{tabular}}
\end{table}

\newpage

\section {The Double Well, Quartic, Anharmonic, Oscillator}
The quartic anharmonic  potential corresponds to an $m_s = 1$ problem, within each parity symmetry class; and is representative of the most general type of problem amenable to OPPQ-BM analysis. 

For the quartic anharmonic oscillator, the ${\cal S}_I(E, {\overrightarrow u})$ functions in Eq.(6) retain their missing moment dependence. We then introduce the purely energy dependent eigenvalue functions, $\lambda_I(E)$, as discussed in Sec. 2. These will allow us to implement the OPPQ-BM eigenenergy bounding formalism.  

We do not have to work with these, $\lambda_I(E)$, functions which correspond to the ${\cal C}({\overrightarrow u}) = u_0^2+u_1^2 = 1$ constraint within the CQFM formulation in Eq.(10). We can implement OPPQ-BM relative to another linear or nonlinear constraint on the missing moments; however, this is not investigated here. This flexibility is the process by which we can extend OPPQ-BM to multidimensional systems, as done in Sec. 5 for the QZM problem. 

All the necessary proofs of OPPQ-BM, for one dimensional formulations using ${\cal L}_I(E) \equiv \lambda_I(E)$ are given here, in the context of the quartic anharmonic oscillator problem.

\subsection{OPPQ-BM: MER and CDR Preliminaries}

For simplicity, we solely focus on the symmetric states of the quartic anharmonic potential, $V(x) = x^4 -5x^2$. We implement the same analysis as that of the harmonic oscillator. The (Stieltjes) moments satisfy the MER representation:
\begin{equation}
u(p+2) = 5u(p+1)+ \ E u(p)\ +\ 2p(2p-1)u(p-1), p \geq 0.
\end{equation}
We note that $\{u(0) \equiv u_0, u(1)\ \equiv u_1 \}$ are the two independent initialization, or {\it missing}, moments. 

The moment equation can be expressed as 
\begin{eqnarray}
u(p) = \sum_{\ell=0}^{m_s=1} M_E(p,\ell) u_{\ell},
\end{eqnarray}
where 
\begin{eqnarray}
 \hspace{-30pt} M_E(p+2,\ell) = 5M_E(p+1,\ell)+ \ E M_E(p,\ell)\ +\ 2p(2p-1)M_E(p-1,\ell), \cr 
\end{eqnarray}
$ p \geq 0$,
and
\begin{eqnarray}
 M_E(\ell_1,\ell_2) = \delta_{\ell_1,\ell_2}, 0 \leq \ell_{1,2} \leq m_s = 1.
\end{eqnarray}

Using the same  CDR representation (and weight) as in the harmonic oscillator case we have (i.e. the Stieltjes representation, for the symmetric states, involves the configuration $\Phi$)
\begin{eqnarray}
c_\eta(E,u_0,u_1) & = & \langle P_\eta|\Phi\rangle,\cr
c_\eta(E,u_0,u_1) & = & \sum_{j=0}^\eta \Xi_j^{(\eta)}u(j),\cr
c_\eta(E,u_0,u_1) & = & \sum_{\ell=0}^{1} \Lambda^{(\eta)}_{E,\ell}\ u_{\ell} = 
 {\overrightarrow{ \Lambda_{E}^{(\eta)}}}\cdot{\overrightarrow u}, 
\end{eqnarray}
where ${\overrightarrow u} \equiv (u_0,u_1)$, and
\begin{eqnarray}
\Lambda^{(\eta)}_{E,\ell} & = & \sum_{j=0}^\eta \Xi_j^{(\eta)}M_E(j,\ell).
\end{eqnarray}

\subsection{The General Form for the OPPQ-BM Quantization Condition}

The partial sums correspond to
\begin{eqnarray}
{\cal S}_I(E, {\overrightarrow u}) & = & \sum_{\eta=0}^I c_\eta^2(E, {\overrightarrow u}),
\end{eqnarray}
or
\begin{eqnarray}
{\cal S}_I(E, {\overrightarrow u}) & = & \sum_{\eta=0}^I ({\overrightarrow\Lambda_{E}^{(\eta)}}\cdot {\overrightarrow u})^2 ,\cr
 & = & \langle{\overrightarrow u}| \sum_{\eta=0}^I 
{\overrightarrow{\Lambda_{E}^{(\eta)}}}{\overrightarrow{\Lambda_{E}^{(\eta)}}}| {\overrightarrow u}\rangle, \cr
 & \equiv & \langle{\overrightarrow u}|{\bf{\cal P}}_I(E)|{\overrightarrow u}\rangle.
\end{eqnarray}
We explicitly identify the indicated matrix in Eq.(56) because it will play an important role in the OPPQ-BM formalism:

\begin{eqnarray}
{\cal P}_I(E) \equiv \sum_{\eta=0}^I 
{\overrightarrow{\Lambda_{E}^{(\eta)}}}{\overrightarrow{\Lambda_{E}^{(\eta)}}} > 0.
\end{eqnarray}
This symmetric matrix is of dimension $1+m_s = 2$. It is positive definite if $I\geq 1$, since it should involve two linearly independent 
${\overrightarrow {\Lambda}}$ vectors. An additional, extremely important property follows from its definition:

\begin{eqnarray}
{\cal P}_{I+1}(E) = {\cal P}_{I}(E) +
{\overrightarrow{\Lambda_{E}^{(I+1)}}}{\overrightarrow{\Lambda_{E}^{(I+1)}}} .
\end{eqnarray}
Thus, if $I \geq 1$, the ${\cal P}_I(E)$ matrices are positive definite, whereas the dyad matrix is semidefinite. More importantly, all the matrices have the same dimension $1+m_s = 2$. 

Define the smallest eigenvalue for the ${\cal P}_I(E)$ matrix:
\begin{eqnarray}
\lambda_I(E) \equiv \ Smallest \ Eigenvalue \ of \ {\cal P}_I(E).
\end{eqnarray}
Since the dimension of the matrices in Eq.(58) are all the same, it follows that
\begin{eqnarray}
0 < \lambda_I(E) < \lambda_{I+1}(E)  < \ldots < \lambda_\infty(E).
\end{eqnarray}
This relation does not hold in the multidimensional case, for the corresponding positive definite matrices, since their dimension changes with the order (i.e. $m_s$) of the expansion.

The energy and missing moment partial sums define a positive, increasing sequence:

\begin{eqnarray}
0 < {\cal S}_{I}(E, \overrightarrow{u}) < {\cal S}_{I+1}(E, \overrightarrow{u}) < \ldots < {\cal S}_\infty(E, \overrightarrow{u}).
\end{eqnarray}
The asymptotic limit, ${\cal S}_\infty(E, \overrightarrow{u})$, corresponds to the expression $\langle \Phi|{1\over {\cal R}}|\Phi\rangle$ which, from Eq.(12), must be finite for physical states, and infinite, for unphysical values for the energy and/or missing moments. This becomes the
 OPPQ-BM quantization condition:

\begin{eqnarray}
Lim_{I\rightarrow \infty}{\cal S}_I(E, {\overrightarrow u}) = \cases{ finite, \iff \ E = E_{phys} \ and \ {\overrightarrow u} = {\overrightarrow u}_{phys}, \cr
\infty, \iff \ E \neq E_{phys} \ or \ {\overrightarrow u} \neq {\overrightarrow u}_{phys}. }
\end{eqnarray}
This is the general type of expression for both one dimensional and multidimensional problems.

\subsection{Solving the OPPQ-BM Quantization Conditions, Eq.(62)}

The challenge is to develop a procedure for solving Eq.(62). As will become clear here and in the next section, for any one dimension, or multidimensional, space problem, the missing moments do not directly contribute towards the quantization process. We can identify purely energy dependent functions whose asymptotic properties mimic those of the ${\cal S}_I(E)$ functions for the harmonic oscillator case (i.e. Eq.(39)); and which will have all the other properties exhibited by these functions, leading to the generation of converging eigenenergy bounds. 

Given the strictly increasing nature of the smallest eigenvalue sequence  in Eq.(60), one may suspect that these expressions will define the desired $S_I(E)$ functions. This will be the case. We first prove certain additional properties of the above eigenvalue sequence. We note that for the one dimensional case, if the focus is on the $\lambda_I(E)$ expressions, since all the missing moment vectors in Eq.(61) are of the same dimension, nothing is lost by using unit missing moment vectors: ${\overrightarrow u} \rightarrow {\hat u}$.

Adapting Eq.(61) to the one dimensional case, involving unit missing moment vectors, we have:
\begin{eqnarray}
Lim_{I\rightarrow \infty}{\cal S}_I(E, {\hat u}) = \cases{ finite, \iff \ E = E_{phys} \ and \ {\hat u} = {\hat u}_{phys}, \cr
\infty, \iff \ E \neq E_{phys} \ or \ {\hat u} \neq {\hat u}_{phys}. }
\end{eqnarray}

The question is: what must be the physical missing moment unit vector, at infinite order,  satisfying Eq.(63), or: 
\begin{eqnarray}
{\cal S}_\infty(E_{phys},{\hat u}_{phys}) = finite.
\end{eqnarray}
 The answer is, it must correspond to the eigenvector with smallest eigenvalue for the 
${\cal P}_{\infty}(E_{phys})$ positive definite matrix defining ${\cal S}_\infty (E,{\hat u}) = \langle{\hat u}|{\cal P}_\infty(E)|{\hat u}\rangle$.

The proof of this important result is by {\it reductio ad absurdum}. Denote by ${\hat u}_\sigma$ the missing moment eigenvector with the smallest eigenvalue, for the 
${\cal P}_\infty(E_{phys})$ matrix:
\begin{eqnarray}
\hspace{-50pt} {\cal S}_\infty(E_{phys},{\hat u}_\sigma) \equiv \lambda_\infty(E_{phys})  \equiv Inf_{\overrightarrow u}\{{\cal S}_{\infty}(E_{phys},{\overrightarrow u})|\ where \ |{\overrightarrow u}|^2 = 1\} . 
\end{eqnarray}
Then if we assume that both unit vectors (i.e. ${\hat u}_{phys}$ and ${\hat u}_{\sigma}$) are different, we obtain a contradiction:
\begin{eqnarray}
{\hat u}_{phys} \neq {\hat u}_\sigma \iff\ Contradiction.
\end{eqnarray}
The simple reason is that the smallest eigenvalue is smaller than the expectation value with respect to any other unit vector. However, since the smallest eigenvalue corresponds to an unphysical (by assumption) missing moment unit vector, it must be infinite, based on the OPPQ-BM quantization condition:
\begin{eqnarray}
\infty = {\cal S}_\infty(E_{phys},{\hat u}_\sigma)   < {\cal S}_\infty(E_{phys},{\hat u}_{phys}) = finite .
\end{eqnarray}
This is the contradiction that validates:
\begin{eqnarray}
\lambda_\infty(E_{phys}) = finite,
\end{eqnarray}
or ${\hat u}_{phys} = {\hat u}_\sigma$.

The previous result strongly suggests that we can replace Eq.(63) with a similar result solely involving the eigenvalue functions:
\begin{eqnarray}
Lim_{I\rightarrow \infty}{\lambda}_I(E) = \cases{ finite, \iff \ E = E_{phys} , \cr
\infty, \iff \ E \neq E_{phys}. }
\end{eqnarray}

Given Eq.(69), combined with Eq.(60), it is now clear that for one dimensional systems the smallest eigenvalue functions,
$ \lambda_I(E)$,
will have exactly the same properties as the $S_I(E)$ functions for the harmonic oscillator. All the properties identified for the harmonic oscillator problem repeat themselves. The  most important are given below.

Define the local minima of $\lambda_{I}(E)$ by $\lambda_{I}(E_I^{(min)})$, where 
\begin{eqnarray}
\partial_E \lambda_{I}(E_I^{(min)}) = 0.
\end{eqnarray}
We then have from Eq. (60):
\begin{eqnarray}
0 < \lambda_I(E_I^{(min)}) <  \lambda_I(E_{I+1}^{(min)}) < \lambda_{I+1}(E_{I+1}^{(min)}),
\end{eqnarray}
as was the case for $S_I(E)$ in Eq. (42). This then generates 
 the positive, increasing, bounded sequence : 
\begin{equation}
0 < \lambda_I(E_I^{(min)}) < \lambda_{I+1}(E_{I+1}^{(min)})<  \ldots < \lambda_{\infty}(E_{phys}) < \infty \ ; \\
\end{equation}
which in turn yield high accuracy estimates for the eigenenergies:
\begin{equation}
Lim_{I\rightarrow \infty}{E_I}^{(min)} = E_{phys}.\\
\end{equation}

Any  coarse upper bound ${\cal B}_U$ to the sequence in Eq.(72), will generate converging bounds to the physical energies.
In particular, the roots of $\lambda_I(E_I^{(L)}) =\lambda_I(E_I^{(U)})  = {\cal B}_U$ define converging lower and uper bounds to the physical energy, 
\begin{eqnarray}
E_{I}^{(L) }< E_{phys} < E_I^{(U)},
\end{eqnarray}
 and
\begin{equation}
Lim_{I\rightarrow \infty}\Big( E_I^{(U)}-E_I^{(L)}\Big) = 0^+.
\end{equation}

\subsection{Numerical Results}

In Tables 2 and 3 we give the OPPQ-BM energy estimates for our problem, and bounds for the fifth even parity state, $E_8$. In Table 3, already at $I=100$ we are confident of a coarse upper bound (i.e. .7) to the $S_I(E_I)$ limit of approximately $.64$. Despite a 9\% coarseness in the ${\cal B}_U$ estimate (i.e. ${{.7-.64}\over {.64}} = 9\%$), we can continue to bound the energy up to $I =250$, achieving a bounding accuracy of $10^{-15}$.

\subsection{Closed form expression for $\partial_E\lambda_I(E)$}

In determining the roots of $\partial_E \lambda_I(E) = 0$, we can exploit the fact that these expressions can be generated in closed form, if the smallest (missing moment) unit eigenvector is generated.  

In terms of its eigenvector, 
$\lambda_I(E_I) = \langle {\hat  u}_I(E_I)|  {\cal P}_I(E_I)|{\hat  u}_I(E_I)\rangle$, we have:

\begin{equation}
\partial_E \lambda_{I}(E_I) =  \langle {\hat u}_{I}(E_I)|{{\partial}\over{\partial E}}{\cal  P}_I(E_I)|{{{\hat u}
_{I}(E_I)}}\rangle = 0.
\end{equation}
The expression ${{\partial}\over{\partial E}}{\cal  P}_I(E_I)$ can be calculated as follows:

\begin{eqnarray}
\partial_E{\cal  P}_I(E) = \partial_E \Big( \sum_{\eta=0}^I{\overrightarrow{\Lambda_{E}^{(\eta)}}}{\overrightarrow{\Lambda_{E}^{(\eta)}}}\Big), \cr
\hspace{48pt}=\sum_{\eta=0}^I{\overrightarrow{\partial_E\Lambda_{E}^{(\eta)}}}{\overrightarrow{\Lambda_{E}^{(\eta)}}} + \sum_{\eta=0}^I{\overrightarrow{\Lambda_{E}^{(\eta)}}}{\overrightarrow{\partial_E\Lambda_{E}^{(\eta)}}};\cr
\end{eqnarray}
and
\begin{eqnarray}
{\overrightarrow{\partial_E\Lambda_{E}^{(\eta)}}} = \partial_E({\Lambda_{E,0}^{(\eta)}},{\Lambda_{E,1}^{(\eta)}},\ldots,{\Lambda_{E,m_s}^{(\eta)}}), \\
{\partial_E\Lambda_{E,\ell}^{(\eta)}} = \sum_{j=0}^\eta\Xi_j^{(\eta)}\partial_E M_E(j,\ell),
\end{eqnarray}
assuming the orthonormal polynomial coefficients are independent of the energy parameter, $E$. Exceptions to this can arise. The expression $\partial_E M_E(j,\ell)$ can be obtained from the moment equation directly, as follows.
\newpage
From Eq.(51) it follows that 
\begin{eqnarray}
 \partial_E M_E(p+2,\ell)   =  5\partial_E M_E(p+1,\ell) + \ E \partial_E M_E(p,\ell)\ + M_E(p,\ell) \cr 
\hspace{100 pt} +\ 2p(2p-1)\partial_E M_E(p-1,\ell), \ p \geq 0, 
\end{eqnarray}
where
\begin{eqnarray}
 \hspace{-10pt} \partial_E M_E(\ell_1,\ell_2) = 0, 0 \leq \ell_{1,2} \leq m_s \ .
\end{eqnarray}
Thus, assuming the $M_E(p,\ell)$ have been generated then the partial derivative with respects to the energy can be also generated.
\\

\begin{table}
\caption{\label{tab1}
OPPQ-BM Energies, $\partial_E \lambda_I(E_I) = 0$,\ $V(x) = x^4 - 5x^2$, $m_s = 1$, $R = e^{-{{x^2}\over 2}}$.}
\centerline{
\begin{tabular}{lccccc}
\hline
$I$ & $E_0$    &  $E_2$  & $E_4$ & $E_6$     &  $E_8$     \\
\hline
 10 & -3.28719572670&  1.19986317656&   9.03942279437&  &   \\
 20& -3.40545008630&     .66975276413&   6.12438920846&  13.8508828043 & \\
 30 &-3.41010592876&     .63936791163&   5.89206898015&  13.5675998541 & 22.6774222840\\
 40 &-3.41014379159&     .63892839165& 5.88534289705&  13.5483205270&  22.6347607487 \\
 50 & -3.41014273834&    .63892037926&     5.88529619878   &       13.5475790455     &      22.6359743625\\
 60 & -3.41014275904&    .63891958477&     5.88529405095  &        13.5475707843       &    22.6363247631\\
70&   -3.41014276124 &   .63891956388 &    5.88529385955   &       13.5475708449     &      22.6363360218\\
80&   -3.41014276124&    .63891956381 &    5.88529385889 &         13.5475708482&           22.6363363374\\
90&   -3.41014276124&    .63891956378 &    5.88529385879&          13.5475708486&           22.6363363803\\
100& -3.41014276124 &   .63891956378 &    5.88529385878&          13.5475708486 &          22.6363363809 \\
 \hline

 \hline
\end{tabular}}
\end{table}

\begin{table}
\caption{\label{tab1}
OPPQ-BM Bounds for $E_8$:  $V(x) = x^4-5x^2$,  $R=e^{-{{x^2}\over2}}$
}
\centerline{
\begin{tabular}{lccll}
\hline
$I$ & $\partial_E \lambda(E^{(min)}_I) = 0$     &  $\lambda_I( E_I)$  & $E_I^{(L)}$     &  $E_I^{(U)}$     \\
\hline
\hline
 30 & 22.6774222840&     .64106446&  21.4017&  23.8979\\
 40 & 22.6347607487&   .64116603& 22.3932& 22.9076 \\
 50 & 22.6359743625&     .64117354&  22.5408&  22.7332\\
 60 & 22.6363247631&     .64117452&  22.6215& 22.6512 \\
 70 & 22.6363360218&  .64117456&  22.6339& 22.6387 \\
 80 & 22.6363363374&  .64117456&  22.6355& 22.6372 \\
 90 & 22.6363363803&  .64117456&  22.636248&       22.636435 \\
 100 & 22.6363363809&  .64117456&22.636304&       22.636368   \\
110 &                    &        &              22.6363308    & 22.6363418             \\
120 &        &      &                            22.6363352    & 22.6363375    \\
130 &   &  &                                     22.6363360    & 22.6363367\\
140 &   &  &                                     22.63633633     & 22.63633643\\
150&    &   &                                    22.6363363640 & 22.6363363970\\
160&    &   & 										  22.6363363789   &22.6363363828  \\
170&     &   &                                    22.6363363800  &22.6363363818 \\
180&       &   &                                  22.63633638079  &22.63633638099 \\
190&       &   &                                  22.63633638084  &22.63633638094 \\
200&       &   &                                  22.636336380885    &22.636336380898 \\
210&       &   &                                  22.6363363808889  &22.6363363808954 \\
220&       &   &                                  22.6363363808908  &22.6363363808928 \\
230&       &   &                                  22.6363363808916  &22.6363363808920 \\
240&       &   &                                  22.636336380891744  &22.636336380891824 \\
250&       &   &                                  22.636336380891776  &22.636336380891798 \\
\hline
  & &  $.64117456< {\cal B}_U = .7$ &  &     \\
\hline
 \hline
\end{tabular}}
\end{table}

\newpage

\section{The Quadratic Zeeman Problem}

In this problem we show the inadequacy of using the $\lambda_I(E)$ functions for implementing OPPQ-BM in multidimensions. Instead, by using different missing moment constraint relations within an ${\cal L}_I(E)$ formulation (i.e. Eq. (10)), we can  implement OPPQ-BM and generate tight bounds, over a broad range of magnetic fields. The necessary proofs that these new energy dependent functions satisfy the basic OPPQ-BM structure is found in the Appendix.

\subsection{The MER Representation}

  For simplicity, we examine  the even parity, zero azimuthal angular momentum states, for the quadratic Zeeman (QZM) problem corresponding to: 
\begin{equation}
\Big( - {1\over 2} \Delta + {{B^2}\over 8} (x^2 + y^2) - {1\over r} - E \Big ) \Psi = 0.
\end{equation}
We adopt the parabolic coordinate representation formalism used by Handy et al [3,4], transforming the three dimensional QZM problem  (atomic units adopted),
into a parabolic coordinate representation defined by $\xi = r-z \geq 0$, $\eta = r+z \geq 0$. Additionally, from EMM we know that a more efficient missing moment structure (i.e. a reduction in the  order of the finite difference equation) is obtained if we transform the wavefunction according to 
\begin{equation} 
\Phi(\xi,\eta) \equiv \Psi(\xi,\eta) exp(-B \xi \eta/4).
\end{equation}
The transformed parabolic partial differential equation becomes
\begin{eqnarray}
\hspace{-50pt} \partial_\xi(\xi\partial_\xi \Phi) + \partial_\eta(\eta\partial_\eta \Phi) + {1\over 2} B\xi\eta(\partial_\xi\Phi+\partial_\eta\Phi) 
+\Big[ {1\over 2}(E + {1\over 2} B) (\xi+\eta) + 1 \Big] \Phi = 0.
\end{eqnarray}
The asymptotic form of the transformed configuration is given by $R(\rho,z) = exp(-{B\over 2}\rho^2) \times exp(-\sqrt{\epsilon\over 2} |z|)$, where $\rho^2=x^2+y^2$, or :
\begin{equation}
\Phi(\xi,\eta) \rightarrow exp\Big[ -{1\over 2}B\xi \eta -({\epsilon\over 2})^{1\over 2}|\eta-\xi|\Big],
\end{equation}
where the binding energy is given by $\epsilon = B/2 - E$.

The two dimensional Stieltjes moments for $\Phi$ are defined by
\begin{equation}
u(m,n) = \int_0^\infty d\xi \ \int_0^\infty d\eta \ \xi^m \eta^n \Phi(\xi,\eta),
\end{equation}
with moment equation
\begin{eqnarray}
\hspace{-50pt} m^2u(m-1,n)+n^2u(m,n-1) \cr 
-{1\over 2}[Bm+\epsilon]u(m,n+1) 
-{1\over 2}[Bn+\epsilon]u(m+1,n)+u(m,n) = 0,
\end{eqnarray}
with even parity invariance ($z \leftrightarrow  -z$ or $\xi \leftrightarrow \eta$) reflected in the moment reflection symmetry $u(m,n) = u(n,m)$. 

The moment equation defines a ``nearest neighbor" pattern in which the $u(m,n)$ moment is linked to the $\{u(m+1,n),u(m-1,n),u(m,n+1),u(m,n-1)\}$ moments, so long as the reflection symmetry is exploited, and the moment indices limited to the nonnegative integers $m,n \geq 0$. 
The missing moments correspond to  $\{u(\ell,\ell) | \ell \geq 0\}$.  For $0 \leq \ell \leq m_s$, the $1+m_s$ missing moments, $u(\ell,\ell) \equiv u_\ell$,  generate all the moments defined through their antidiagonal index: $\{u(m,n)| m+n \leq 2m_s+1\}$. In this manner we generate the moment - missing moment relation:
\begin{equation}
u(m,n) = \sum_{\ell = 0}^{m_s} M_\epsilon(m,n,\ell) u_\ell, \ where \ 0\leq m+n \leq 2m_s+1,
\end{equation}
$u_\ell \equiv u(\ell,\ell)$ and $M_\epsilon(\ell_1,\ell_1,\ell_2) = \delta_{\ell_1,\ell_2}$.  

Given the first $1+m_s$ missing moments, $\{u_\ell |0 \leq \ell \leq m_s\}$,  a finite number of moments are generated, defining the ${\cal U}_{m_s}$ subspace. These subspaces form a nested hierarchy, ${\cal U}_{m_s} \subset {\cal U}_{m_s+1}\subset \ldots \subset {\cal U}_\infty$. The MER relation is an exact projection of the Schrodinger equation into each of these subspaces. 

The binding energy matrix coefficients, $M_\epsilon(m,n,\ell)$, satisfy the moment equation with respect to the $(m,n)$ indices and the given initialization conditions:
\newpage
\begin{eqnarray}
\hspace{-70pt} m^2M_\epsilon(m-1,n,\ell)+n^2M_\epsilon(m,n-1,\ell)  -{1\over 2}[Bm+\epsilon]M_\epsilon(m,n+1,\ell) \cr 
\hspace{70pt} -{1\over 2}[Bn+\epsilon]M_\epsilon(m+1,n,\ell)+M_\epsilon(m,n,\ell) = 0,
\end{eqnarray}
where $M_\epsilon(\ell_1,\ell_1,\ell_2) = \delta_{\ell_1,\ell_2}$.

\subsection{Generating the Orthonormal Polynomials}

The preferred  {\it reference function - weight} is any expression which takes on the  asymptotic form of the physical solutions. Instead of using the expression in Eq.(85), an easier expression to use (with respect to generating the required power moments of the weight, and in turn the coefficients of the orthonormal polynomials) is $R_{QZM}(\rho,r) = exp(-{B\over 2}\rho^2) \times exp(-\sqrt{\epsilon\over 2} r)$:

\begin{equation}
R_{QZM}(\xi,\eta) = exp\Big( -{1\over 2}B\xi\eta - ({\epsilon\over 2})^{1\over 2} (\xi+\eta)\Big),
\end{equation}
with  power moments 
\begin{equation}
w_{QZM}(m,n) =  \int_0^\infty d\xi \int_{0}^\infty d\eta \  \xi^m \eta^n  exp\Big( -\beta\xi\eta - \alpha (\xi+\eta)\Big),  \nonumber 
\end{equation}
\begin{equation}
\hspace{65pt} \equiv {{n!}\over{\alpha^{m+n+2}}}\Omega(m,n+1,g)
\end{equation}
where $\alpha = ({\epsilon\over 2})^{1\over 2} $, $\beta = {1\over 2}B$, and $g = {\beta \over{\alpha^2}} = {B\over \epsilon}$. The $\Omega$ functions are recursively generated as follows. First, $\Omega(0,1,g) < 1$ is numerically determined to high accuracy. This then allows us to generate
\begin{equation}
\Omega(0,n+1,g) = \sum_{j=1}^n {{(-1)^{j+1}}\over {g^j}}{{(n-j)!}\over{n!}}  +{{(-1)^{n}}\over{g^nn!}}\Omega(0,1,g),
\end{equation}
for $n \leq N$. For each such `$n$', we can generate 
\begin{equation}
\hspace{-70pt} \Omega(m+1,n+1,g) = {1\over g} \delta_{m,0} + {m\over g} \Omega(m-1,n+1,g) + [m-n-g^{-1}] \Omega(m,n+1,g), 
\end{equation} 
for $0 \leq m \leq M$.

One can allow the reference function to incorporate the binding energy parameter, as given above. This makes the generation of the orthonormal polynomials more time consuming. We do this to low order to obtain an estimate of the physical binding energy (i.e. $\epsilon_0 \approx \epsilon_{phys}$). Once this is determined, we then keep $\epsilon_0$ fixed within $R_{QZM}$, and keep $\epsilon$ as a variable within the moment equation. So long as $\epsilon > \epsilon_0$, we preserve the asymptotic requirements of the OPPQ formalism. Implementing the above process for $\epsilon_0$, we find that it corresponds to the first significant figure for the (eventual) physical energy. The data in Tables 4 are generated on this basis.

The orthonormal polynomials will take on the form
\begin{eqnarray}
P_I(\xi,\eta) \equiv \sum_{j=0}^I\Xi_j^{(I)} \xi^{m_j} \eta^{n_j},
\end{eqnarray}
for some appropriate coordinate pair sequence ordering, $\{(m_j,n_j)|j = 0,1,2,\ldots\}$. This sequence ordering must map into the set of nonnegative coordinate integer pairs in a one-to-one and onto manner. Given that the missing moments generate the previously identified moment subspaces, ${\cal U}_{m_s}$, the most efficient sequence ordering must emulate this as well. The most natural choice is in a progression based on their antidiagonal sum: $(0,0)_0, (1,0)_1, (0,1)_2, (2,0)_3, (1,1)_4, (0,2)_5,\ldots$.

The orthonormal polynomials must satisfy the orthonormal relations relative to the chosen weight, $\langle {\overrightarrow \Xi^{(I)}}|{\cal W}|{\overrightarrow \Xi^{(J)}}\rangle = \delta_{I,J}$, where the positive Hankel moment matrix is given by
${\cal W}_{i,j}\ \equiv w_{QZM}(m_i+m_j,n_i+n_j)$. The coefficients are then obtained through the  Cholesky decomposition  ${\cal W} = {\cal C}{\cal C}^\dagger$, resulting in ${\overrightarrow \Xi^{(I)} }= \Big({\cal C}^\dagger \Big)^{-1}{\hat e}_I$, where ${\hat e}_I$ is the unit coordinate vector in the $I$-th direction. 

\subsection {The CDR-MER/OPPQ Representation}

Assembling all the OPPQ-BM components we have the following. 
The CDR-MER/OPPQ expansion takes on the form
\begin{eqnarray}
\Phi(\xi,\eta) = \sum_{I=0}^\infty c_I \ P_I(\xi,\eta) \ R_{QZM}(\xi,\eta),
\end{eqnarray}
and the projection coefficients become (i.e. $c_I = \langle P_I|\Phi\rangle$) 
\begin{equation}
c_I = \sum_{j=0}^I \Xi_j^{(I)} u(m_j,n_j),
\end{equation}
or
\begin{equation}
c_I(\epsilon,{\overrightarrow u}) = \sum_{\ell=0}^{m_s(I)}\Lambda^{(I)}_{\epsilon,\ell}u_\ell,
\end{equation}
where $m_s(I)$ is the missing moment order required to generate $c_I$, and 
\begin{equation}
\Lambda^{(I)}_{\epsilon,\ell} = \sum_{j=0}^I\Xi_j^{(I)}M_\epsilon(m_j,n_j,\ell).
\end{equation}

An alternative way to use the above relations is to say that  the first $1+m_s$ missing moments, $\{u_\ell| 0 \leq \ell \leq m_s\}$, can be used to generate all the moments $\{u(m,n)| 0\leq m+n \leq 2m_s+1\}$, through the moment-missing moment relation in Eq. (87). However, these are the moments required in order to generate all the sequentially ordered $c_I$ coefficients satisfying $\{c_I | 0 \leq I \leq  I_{m_s} \equiv (m_s+1)(2m_s+3)-1\}$ in Eq. (98). These $c_I$ coefficients depend on the coefficients of the orthonormal polynomials for the same range of $I$-index values. However, these coefficients require a Cholesky analysis relative to the $R_{QZM}$-moment matrix ${\cal W}_{i,j} =w_{QZM}(m_i+m_j,n_i+n_j)$ where $m_i+m_j+n_i+n_j \leq 2(2m_s+1)$, and $0 \leq I \leq I_{m_s}$. That is, the generation of the $\Omega$'s requires $M+N \leq 2(2m_s+1)$, as defined through Eqs.(93,94). 

The corresponding partial sums, ${\cal S}_I$, become :
\begin{eqnarray}
{\cal S}_I(\epsilon,{\overrightarrow u}) & =  &\sum_{i=0}^I\Big( c_i(\epsilon , {\overrightarrow u})\Big)^2, \cr
& = & \sum_{\ell_1 = 0}^{m_s(I)}\sum_{ \ell_2= 0}^{m_s(I)} u_{\ell_1}{\cal P}_{I;\ell_1,\ell_2}(\epsilon)   u_{\ell_2} ,\cr
{\cal P}_{I;\ell_1,\ell_2}(\epsilon) & \equiv & \sum_{i=0}^I \Lambda_{\epsilon,\ell_1}^{(i)} \Lambda_{\epsilon,\ell_2}^{(i)},\cr
{\cal P}_I({\epsilon}) & = & \sum_{i=0}^I {\overrightarrow {\Lambda_\epsilon^{(i)}}}{\overrightarrow {\Lambda_\epsilon^{(i)}}},
\end{eqnarray}
involving a   symmetric positive definite matrix, ${\cal P}_I(\epsilon)$, made up of indiviual semidefinite dyadic matrices. This matrix, ${\cal P}_I(\epsilon)$, is positive definite because `$I$' is usually much larger than the dimension, $1+m_s$, of the $\Lambda$-vectors. The `$I$' index determines the number of missing moments, $m_s(I)$, required.  

We can think of ${\overrightarrow u}$ as an infinite dimensional missing moment vector, of which only the first $1+m_s(I)$ components, ${\overrightarrow u} \rightarrow (u_0,u_1,\ldots,u_{m_s(I)})$, contribute to the ${\cal S}_I$ quadratic form. This perspective is important when dealing with the increasing, strictly positive sequence,
\begin{eqnarray}
0 < {\cal S}_I(\epsilon,{\overrightarrow u}) < {\cal S}_{I+1}(\epsilon,{\overrightarrow u}) <  \ldots < {\cal S}_\infty(\epsilon,{\overrightarrow u}), \nonumber
\end{eqnarray}
 generated at fixed $\epsilon$ and for a fixed, infinite dimensional missing moment vector: ${\overrightarrow u} = (u_0,u_1,\ldots,u_\infty)$.
In this sequence progression, the number of missing moments will stay fixed at $1+m_s$, for $I_{m_s-1}+1\leq I \leq I_{m_s}$, after which it will increase by 1 to $2+m_s$, etc. Due to this change in dimensionality, the missing moment constraint in Eq.(10), corresponding to the constrained quadratic form minimization (CQFM) analysis, must be chosen in some uniform manner, starting at some minimal ``$I$" value. As noted earlier, for one dimensional problems, the number of missing moments is fixed; whereas for multidimensional problems, it changes with the order of the OPPQ-BM analysis.We further elaborate on this CQFM procedure below and in the Appendix.

As noted, all the orthonormal polynomials with index $I$ satisfying $I \leq I_{m_s}$ as defined previously, will depend on the first $1+m_s$ missing moments.  Those with index greater than this, $I_{m_s}+1 \leq I \leq I_{m_s+1}$, will depend on the first $2+m_s$ missing moments. It is at the transition point $I=I_{m_s} \rightarrow I=I_{m_s}+1$ that the dimensionality of the positive definite matrices changes. We make this explicit:
\begin{eqnarray}
{\cal P}_{I_{m_s} +1}(\epsilon) = {\cal P}_{I_{m_s} }(\epsilon) + {\overrightarrow {\Lambda_\epsilon^{(I_{m_s})}}}{\overrightarrow {\Lambda_\epsilon^{(I_{m_s})}}}.
\end{eqnarray}
This relation involves two positive definite matrices and one semidefinite (dyadic) matrix. The dimensions of each satisfy:
 $Dim\big( {\cal P}_{I_{m_s} +1}\big) =  Dim\big({\overrightarrow {\Lambda_\epsilon^{(I_{m_s})}}}{\overrightarrow {\Lambda_\epsilon^{(I_{m_s})}}}\big)$ and $Dim\big( {\cal P}_{I_{m_s} +1}\big)=Dim\big( {\cal P}_{I_{m_s} }\big)+1$. Due to this, one cannot conclude any relationship between the eigenvalues of the two positive definite matrices, as was the case for one dimensional problems (Eq.(9)), where all positive definite ${\cal P}$ matrices have the same dimension. Due to this change in dimensionality,  the more general CQFM analysis (i.e. Eq.(10), where the expansion order ``$n$" is replaced by $m_s$), with a different missing moment vector constraint normalization, is required. That is, the standard normalization $\sum_{\ell =0}^{m_s} u^2_{\ell} =1$ cannot be applied consistently across all ${\cal U}_{m_s}$, for $m_s = 0,1,\dots$, subspaces generated by the missing moments. The Appendix further elaborates on this.

\subsection{Implementation of the Constrained Quadratic Form Minimization}

We will be adopting  the normalization condition 
\begin{eqnarray}
u_0 \equiv 1,
\end{eqnarray}
which is expected to be valid (i.e. not interfere with any symmetry conditions) for the even parity states. Such a normalization is possible from the physics perspective; and is mathematically valid within our OPPQ framework. 

The uniform, linear, normalization, $u_0 = 1$, leads to a constrained quadratic form ${\cal S}_I(\epsilon,{\overrightarrow u}) = {\cal S}_I(\epsilon,(u_0 =1,u_1,\ldots,u_{m_s}))$ whose global minimum value over the unconstrained missing moment variables (i.e. $\{u_\ell | 1\leq \ell \leq m_s\}$)  defines the energy dependent function, ${\cal L}_I(\epsilon)$, introduced in Sec. 2 and further studied in the Appendix.

\begin{eqnarray}
\hspace{-5pt} {\cal S}_I(\epsilon, {\overrightarrow u})   = \sum_{\ell_1 = 0}^{m_s}\sum_{\ell_2=0}^{m_s } u_{\ell_1} {\cal P}_{I;\ell_1,\ell_2}(\epsilon) u_{\ell_2}, \cr
\hspace{-70pt}  {\cal S}_I(\epsilon, (1,u_1,\ldots,u_{m_s}))  =  {\cal P}_{I;0,0}(\epsilon)+ 2 \sum_{\ell=1}^{m_s }  {\cal P}_{I;0,\ell}(\epsilon) u_{\ell} + \sum_{\ell_1 = 1}^{m_s}\sum_{\ell_2=1}^{m_s } u_{\ell_1} {\cal P}_{I;\ell_1,\ell_2}(\epsilon)u_{\ell_2} , \cr
 \hspace{42pt} \equiv  C_I(\epsilon)+ 2{\overrightarrow B}_I(\epsilon)\cdot {\overrightarrow u} + \langle {\overrightarrow u}|{\bf A}_I(\epsilon)|{\overrightarrow u}\rangle.
\end{eqnarray}
 The definitions for the $C_I, {\overrightarrow B}_I,\ and \  {\bf A}_I $ are self-evident by association. 

The global minimum, for fixed $\epsilon$, 
\begin{eqnarray}
{\cal L}_I(\epsilon) = Inf_{\overrightarrow u}\{{\cal S}_I(\epsilon, {\overrightarrow u})| u_0 = 1\},
\end{eqnarray}
corresponds to the solution
\begin{eqnarray}
{\overrightarrow u}_{I;opt} (\epsilon) = - {({\bf A}_I(\epsilon))}^{-1}{\overrightarrow B}_I(\epsilon),
\end{eqnarray}
yielding
\begin{eqnarray}
 {\cal L}_{I}(\epsilon) = C_I(\epsilon) - \langle {\overrightarrow B}_I(\epsilon)|{{\bf A}_I^{-1}}(\epsilon)|{\overrightarrow B}_I(\epsilon)\rangle.
\end{eqnarray}
Clearly, ${\overrightarrow u}_{I;opt}(\epsilon)$, a non-unit vector, is the counterpart to the ``eigenvector of smallest eigenvalue" within the one dimensional   formulation;  whereas ${\cal L}_I(\epsilon)$ is the counterpart to $\lambda_I(\epsilon)$ within the one dimensional formulation, as well.  

As outlined in the Appendix, ${\cal L}_I(\epsilon)$ has all the properties associated with $\lambda_I(E)$ in the one dimensional case (i.e. Sec. 4, for the quartic anharmonic oscillator case; and $S_I(E) = \lambda_I(E)$ for the harmonic oscillator in Sec. 3). That is, the corresponding positive, increasing, sequences can be generated, and from this, eigenenergy estimates and bounds  produced, in the exact manner as in the harmonic oscillator and quartic anharmonic oscillator examples.

\subsection {QZM Numerical Results}

 Table 4 summarizes the  OPPQ-BM results for QZM, including the energy estimates (column three), $\epsilon_{I_{m_s}}^{(min)}$, defined by the local minima relations $\partial_{\epsilon} {\cal L}_{I_{m_s}}(\epsilon_{I_{m_s}}^{(min)}) = 0$; and the energy bounds (columns four and five),  based on a constrained minimization analysis of the quadratic form given in Eqs.(102 - 106).  The ground and first excited states within the even parity, zero azimuthal angular momentum, symmetry class correspond to $\epsilon_{gr,1}$, respectively. The sixth column is the $\epsilon_0$ parameter value used for the reference function weight, as explained earlier. We emphasize that the bounds  are true bounds for the physical energies. The actual positive sequences, and their convergence behavior, are not given here due to space limitations, but may be found in Ref.[18].

For comparative purposes, we quote the energy estimates reported by  Kravchenko et al [6], which appear to be the more accurate estimates in the literature,  yielding twelve-thirteen significant figures for the ground state binding energy, $\epsilon_{gr}$, for magnetic field values $B \leq 4000$. Their results for the first excited state, $\epsilon_1$, vary from twelve significant figures to six, for magnetic field strengths $B \leq 1000$, with no energies reported for higher magnetic fields.  The OPPQ-BM estimates in Table 4 exceed or match their reported $\epsilon_{gr}$ values provided $B \leq 200$.  For $B = 2000$, the OPPQ-BM results for $\epsilon_{gr}$  generate approximately nine of the thirteen significant figures. The only limitation of OPPQ-BM is the computational speed of our computing platform (i.e. MacBook Pro 2.2 GHz/1333MHz). 

For the first excited state, $\epsilon_1$, OPPQ-BM matches or surpasses the reported accuracy of Kravchenko et al's [6] results  for $B \leq O(200)$.  For $B = 2000$, the OPPQ-BM results for $\epsilon_1$ are compared to those of Schimerczek and Wunner [7]; while for $B= 10^4$ we also compare both states to their B-spline analysis results. The ground state results manifest faster convergence than the first excited state.  The generated OPPQ-BM bounds are modest, at these higher magnetic field strengths, given the higher expansion orders required for implementing OPPQ-BM. Tighter bounds would be generated on a faster computer platform, or through an alternate choice to the MER representation chosen here. These possibilities are currently under investigation.

For large magnetic fields, the expansion order required  to obtain results comparable to those in the literature increases. Results corresponding to $I_{m_s} > O(40)$ requires considerable time (i.e. several hours), with available computing resources.

\begin{table}
\caption{ OPPQ-BM  Estimates  and Bounds for QZM: $ \{+,l_z = 0\}$  }
\centerline{
\begin{tabular}{rllllc}
\hline
$B$ & $m_s$ & $\partial_\epsilon{\cal L}_{I_{m_s}}(\epsilon_{I_{m_s}}^{(min)}) = 0$  &    {\it Lower Bound }  & {\it Upper Bound }  & $\epsilon_0$    \\
\hline
\hline
0.02 & 22 & ${0.509900044089401317}_{gr}$ &   0.509900044089401316 &   0.509900044089401318 &0.5 \\
        &      &$0.509900044089\ [6]$ &                                                        &                                         &                                     \\
\hline
        &   22   &  $0.13362417753479289364_1$      &  $0.13362417753479289$  &0.13362417753479291    &0.1     \\
       &      &  $0.133624177534\ [6]$      &               &    \\
\hline
0.20 & 20 & $0.59038156503476258477_{gr}$ & $0.59038156503476258474$  & $0.59038156503476258480$ &0.5\\
      &        &    $0.590381565035\ [6]$                                           &                                    &                              &                                             \\
\hline
      &  28    &    $0.14898667819813574696_{1}$&      0.14898667819813574694& 0.14898667819813574698  & 0.1 \\
        &        & $0.148986678198\ [6]$           &                           &                         &       \\
\hline
 2 & 20 &  $1.02221390766512912_{gr}$ &1.02221390766512894   &1.02221390766512930 &1.0 \\
    & 		&$1.022213907665\ [6]$ 			&			 &	 							 				&		\\
\hline
 & 34&     $0.1739447059728_1$ &0.1739447059 & 0.1739447069 &0.1  \\
 &		&		0.173944705973 [6]   &                           &                        &                                                      \\
\hline
20 & 24 &  $2.21539851543322_{gr}$ &  $2.2153985154326$ &  $2.2153985154375$ &2.0 \\
  &  &  2.215398515433 [6]&   &   &\\
\hline
  & 44&     $0.22384212729_1$ & 0.223842118 &  0.223842138 &0.2  \\
  &    &     0.223842127 [6]    &  &            &  \\
\hline
200& 44 &  $4.72714511068704_{gr}$ & 4.727145110662  &  4.727145110700 &4.0 \\
   &    &      $4.727145110687 [6]$ &  & \\
\hline
&50&        $0.2689772_1$                                & 0.26895            &   .26920     &0.2     \\
 & &            0.2689682 [6] & &       & \\
\hline
2000 & 46 &  $9.304765094_{gr}$ & 9.30475796875&  9.30476699219 &9.0  \\
         &   & $9.304765082770\ [6]$  &     &   \\
\hline
  &    40       &  $0.313_1$                        &             & & 0.3\\
 & 48 & $0.3091_1$ &                                            & &      \\
   &    &0.30624125 [7] &         &        \\
 \hline
10000 & 40 & $14.140995_{gr} $ & 14.137    & 14.143 &14.0\\
&44&14.1409812 &   &   &     \\
&50 &14.1409730 &  &   &  \\
           &         &     $14.14096855\ [7]$                     &               &                                                                \\
\hline
 & 32 &   $0.39533_1 $          &         & &0.3\\
& 40 & 0.37289 &   & &\\
   &    &0.3277107 [7] &         &       & \\
\hline
\end{tabular}}
\end{table}
\newpage
\section{Conclusion}
We have shown that for low dimension Schrodinger equation problems that can be transformed into a moment equation representation (MER), the Christoffel-Darboux expansion in terms of the orthonormal polynomials of the weight, can generate eigenenergy bounds and high accuracy eigenenergy estimates, through a simple algebraic procedure. We have demonstrated its capabilites through several one and two dimensional problems, including the important quadratic Zeeman interaction. This method, OPPQ-BM, can be applied to both bosonic and fermionic systems, and expands the computational tools available to researchers. Given that there are many important physical systems admitting MER representations, the approach presented should be of great interest to many. One might characterize OPPQ-BM as an algebraic {\it shooting method}, applicable to multidimensions, since the ${\cal B}_U$ parameter is empirically determined. This connotation is acceptable so long as it is understood that unlike traditional shooting methods, with a `` hit or miss" philosophy, the OPPQ-BM formalism is a well defined,  controlled, procedure for accurately determining bounds on a desired eigenenergy.

\section{Acknowledgement}
The author is appreciative of Dr. John R. Klauder and Dr. Daniel Bessis for inspiring comments and insight received over many decades that led to the realization of this work. The technical assistance of Dr. Maribel Handy is greatly appreciated.


\newpage
\section{ Appendix: Additional Comments and Proofs for the OPPQ-BM Formalism}

All the main elements of the OPPQ-BM formalism have been presented and argued, except for several. The more important of these is to show that the constrained quadratic form minimization (CQFM) formalism yields an acceptable alternative to the $\lambda_I(E)$ eigenvalue functions. That is, the expressions ${\cal L}_I(E)$, defined through Eq.(10), and its multidimensional generalization, as represented by Eqs.(102-106),  will have all the properties of the $S_I(E) \equiv \lambda_I(E)$ for the harmonic oscillator problem (i.e. Secs. 3), or  equivalently, that of the $\lambda_I(E)$ functions for the quartic anharmonic oscillator, as developed in Sec. 4. Before arguing  this, we discuss, briefly why the eigenvalue relation in Eq.(9) fails for the multidimensional case.

\subsection{Unsuitability of $\lambda_I(E)$ for Multidimensional OPPQ-BM Implementation}

In the mutlidimensional case, the  positive definite matrices given in Eq.(101) will have dimensions that increase with the order of the OPPQ-BM expansion (i.e. the missing moment order, $m_s$). In such cases, the smallest eigenvalues do not satisfy the desirable properties of one dimensional systems, as given in Eq.(9), or Eq.(60). Such relations, combined with the asymptotic properties given in Eq.(69), are central to the eigenenergy bounding capabilities of OPPQ-BM. By introducing the Constrained Quadratic Form Minimization (CQFM) formalism represented in Eq.(10) or Eqs. (101-106), we can enlarge the class of purely energy dependent functions that permit the extension of the OPPQ-BM philosophy to multidimensions. To further clarify this important point, we outline the failure of Eq.(9) for multidimensional systems, below.

As in Eq.(101), assume ${\bf D}_1 = {\bf D}_0+ {\bf S}$, where all matrices  are real and symmetric, ${\bf D}_{0,1} $ are positive definite, while ${\bf S}$ is semidefinite. Assume $Dim({\bf D}_1) = Dim({\bf S}) >  Dim({\bf D}_0)$, where the last row(s) and column(s) of ${\bf D}_0$ are zero. The matrix
 ${\bf D}_0$ is not a principal submatrix of ${\bf D}_1$, hence the eigenvalue interlacing property does not apply.

Let ${\hat{\cal E}}_1$ denote  the smallest eigenvalue eigenvector for ${\bf D}_1$, with $\lambda_{0,1}$ the respective eigenvalues for the positive matrices. Let 
${\overrightarrow {\cal E}}_{1;p}$ denote the vector that results from setting to zero the last component(s) of the corresponding eigenvector; with  $|{\overrightarrow {\cal E}}_{1;p}| < 1$. We then have
\begin{eqnarray}
\lambda_1 \equiv \langle {\hat{\cal E}}_1|{\bf D}_1|{\hat{\cal E}}_1\rangle & = & \langle {\hat{\cal E}}_1|{\bf D}_0|{\hat{\cal E}}_1\rangle+\langle {\hat{\cal E}}_1|{\bf S}|{\hat{\cal E}}_1\rangle,  \cr
& = & \langle {\overrightarrow{\cal E}}_{1;p}|{\bf D}_0|{\overrightarrow{\cal E}}_{1;p}\rangle+\langle {\hat{\cal E}}_1|{\bf S}|{\hat{\cal E}}_1\rangle, \cr
& = & |{\overrightarrow {\cal E}}_{1;p}|^2 {{\langle{\overrightarrow{\cal E}}_{1;p}|{\bf D}_0|{\overrightarrow{\cal E}}_{1;p}\rangle}\over{ |{\overrightarrow {\cal E}}_{1;p}|^2  }}+\langle {\hat{\cal E}}_1|{\bf S}|{\hat{\cal E}}_1\rangle,\cr
& > &|{\overrightarrow {\cal E}}_{1;p}|^2 \lambda_0.
\end{eqnarray}
Since $|{\overrightarrow {\cal E}}_{1;p}|^2 < 1$, nothing can be concluded regarding the relative magnitudes of the eigenvalues. However, we can conclude $\lambda_1 \geq \lambda_S$, the latter being the eigenvalue of the ${\bf S}$, even if it were positive definite.

\subsection{The Relevance of Constrained Quadratic Form Minimization for OPPQ-BM Extension to Multidimensions}

In order to extend OPPQ-BM to multidimensions we must consider a larger class of missing moment constraints. Thus, we focus on:
\\
 \begin{eqnarray}
\hspace{-40pt} {\cal L}_{I}(E) \equiv  Inf_{\overrightarrow \mu}\{{\cal S}_{I}(E,{\overrightarrow \mu}) | {\cal C}({\overrightarrow \mu})  = 1\},
\end{eqnarray}
where the constraint is arbitrary. Clearly, for ${\cal C}({\overrightarrow \mu}) = |\overrightarrow \mu|^2 =1$, then ${\cal L}_I(E) = \lambda_I(E)$. 

Instead of the nonlinear unit vector constraint for the missing moment vector, we can impose other nonlinear,  or linear, normalizations which only constrain a subset of the components of the missing moment vector. For the QZM states of interest (even parity and $L_z = 0$), we can simply impose the  linear normalization constraint $\mu_0 =1$, which will not filter out any targeted physical states due to symmetry requirements.    Other normalization conditions are possible (i.e. $\mu_0^2+\mu_1^2 =1$, etc.), but not considered here.

 The reason we only want to focus on a finite number of the missing moment vector components is that although Eq.(61) is valid in multidimensions, as $I$ increases, more missing moment vector components contribute to the elements of the sequence. For one dimensional systems, the ${\overrightarrow u}$ vector is of fixed dimension; however, for multidimensional systems, this same sequence relation involves more and more components. Therefore, constraints must involve the same components across all ${\cal U}_{m_s}$ subspaces being considered. This is not possible if we insist on unit normalizations for the missing moment vectors.

If we adopt the normalization $\mu_0 = 1$, we then obtain the results in Eqs(102-106). Our interest is to show that for the adopted normalization, and other possible normalizations, the two essential components of the OPPQ-BM formalism, identified below, are preserved. These correspond to the relations given in Eq.(14) and Eq.(11), in that order. We show these to be true.

\subsection{ Proof of Eq.(14)}
First, we must prove that:
\begin{eqnarray}
{\cal L}_\infty(E) = \cases{finite, \iff E = E_{phys} \cr \infty, \iff E \neq E_{phys}\cr}.
\end{eqnarray}
The proof  is also by reductio ad absurdum and similar to that used for Eq.(69), for the quartic anharmonic oscillator example in which ${\cal L}_I(E) \equiv \lambda_I(E)$.  

Define ${\overrightarrow \mu_{\sigma;{\cal C}}}$ as the optimal solution to the constrained quadratic form minimization problem at infinite order, for the physical energy:
\begin{eqnarray}
 {\cal S}_\infty(E_{phys},{\overrightarrow \mu_{\sigma;{\cal C}}}) \equiv {\cal L}_\infty(E_{phys})  = Inf_{\overrightarrow \mu}\{{\cal S}_{\infty}(E_{phys},{\overrightarrow \mu}) | {\cal C}({\overrightarrow \mu})  = 1\};
\end{eqnarray}
whereas the missing moment vector ${\overrightarrow\mu}_{phys;{\cal C}}$ satisfies

\begin{eqnarray}
{\cal S}_\infty(E_{phys},{\overrightarrow \mu}_{phys;{\cal C}}) = finite.
\end{eqnarray}
\subsubsection{Reductio Ad Absurdum}:   

\begin{eqnarray}
{\overrightarrow \mu}_{phys;{\cal C}} \neq {\overrightarrow \mu}_{\sigma:{\cal C}} \iff \ Contradiction.
\end{eqnarray}

(Proof) From the infimum property: ${\cal L}_{\infty}(E_{phys}) =  {\cal S}_{\infty}(E_{phys},{\overrightarrow \mu}_{\sigma;{\cal C}}) \leq {\cal S}_{\infty}(E_{phys},{\overrightarrow \mu}_{phys;{\cal C}}) = finite$. However, the fact that 
${\overrightarrow\mu}_{\sigma;{\cal C}}$ is not a physical vector means that ${\cal S}_\infty(E_{phys},{\overrightarrow \mu}_{\sigma;{\cal C}})  = \infty$. This leads to a contradiction: 
$ \infty={\cal S}_\infty(E_{phys},{\overrightarrow \mu}_{\sigma;{\cal C}}) \leq {\cal S}_\infty(E,{\overrightarrow \mu}_{phys;{\cal C}}) = finite$. Thus we must conclude that 
\begin{eqnarray}
{\overrightarrow \mu}_{phys;{\cal C}} = {\overrightarrow \mu}_{\sigma},
\end{eqnarray}
 and Eq.(109) is valid.

\subsection{Proof of Eq.(11)}

The next requirement is proving that 
\begin{eqnarray}
{\cal L}_I(E) < {\cal L}_{I+1}(E).
\end{eqnarray}
The following analysis uses $I_{m_s}$ and not $I$. The reason is that ``$I+1$" represents the maximum number of orthonormal polynomials generated.  If $I \leq I_{m_s}$ the CDR projection coefficients will involve all the first $1+m_s$ missing moments. Thus, if we take $I=I_{m_s}$ and $I = I_{m_s+1}$, we are working with $1+m_s$ and $2+m_s$ missing moments respectively, and our proof below correctly takes into account the changing dimension  of the associated ${\cal P}_{I_{m_s}}$ positive definite matrices. 

The validity of Eq.(114)  follows from the fact that in general, the multidimensional version of the relation ${\cal S}_{I_{m_s}}(E,{\overrightarrow \mu}) < {\cal S}_{I_{m_s+1}}(E,{\overrightarrow \mu})$, from Eq.(61), will involve increasingly more components of the same (infinite dimensional) missing moment vector. Thus, these relations actually correspond to expressions of the type  ${\cal S}_{I_{m_s}}(E,(\mu_0,\mu_1,\ldots,\mu_{m_s})) < {\cal S}_{I_{m_s+1}}(E,(\mu_0,\mu_1,\ldots,\mu_{m_s},\mu_{m_s+1})) < \ldots$.

Let ${\overrightarrow \mu}_{m_s} \equiv (1,{\overrightarrow u})$, where ${\overrightarrow u}$ corresponds to all the $1 \leq \ell \leq m_s$ components, after the ${\cal C}({\overrightarrow \mu}) = u_0 = 1$ constraint (or any other appropriate  constraint) is imposed. Also ${\overrightarrow \mu}_{m_s; opt}(E)$ refers to the optimal solution for  ${\cal L}_{I_{m_s}}(E) = {\cal S}_{I_{m_s}}(E,{\overrightarrow \mu}_{m_s;opt}(E)) = Inf_{\overrightarrow \mu}\{{\cal S}_{I_{m_s}}(E,{\overrightarrow \mu})|{\cal C}({\overrightarrow \mu}) = 1\}$. Note that the energy,$E$, is arbitrary and fixed. We can make implicit the  $E$ dependence of the optimal missing vector solution, since this is understood.

We then obtain:
\begin{eqnarray}
 {\cal S}_{I_{m_s}}(E,1,{\overrightarrow u}_{m_s+1;opt}) <
{\cal S}_{I_{m_s+1}}(E,1,{\overrightarrow u}_{m_s+1;opt}) = {\cal L}_{m_s+1}(E) .\cr
\end{eqnarray} 
This follows from the general relation in Eq.(61) as applied to multidimensional systems. Furthermore, the expression on the left will only involve the $1 \leq \ell \leq m_s$ components of the ${\overrightarrow u}_{m_s+1;opt}$ vector (which is of dimension $m_s+1$). However, this expression is an upper bound to the optimal solution generated for ${\cal L}_{I_{m_s}}(E,{\overrightarrow \mu})$, as expressed by:
\newpage
\begin{eqnarray}
\hspace{-50pt}{\cal L}_{I_{m_s}}(E) \equiv {\cal S}_{I_{m_s}}(E,(1,{\overrightarrow u}_{{{m_s};opt}})  ) < {\cal S}_{I_{m_s}}(E,1,{\overrightarrow u}_{m_s+1;opt})  \cr
\hspace{170pt}<{\cal S}_{I_{m_s+1}}(E,1,{\overrightarrow u}_{m_s+1;opt}) = {\cal L}_{m_s+1}(E) .\cr
\end{eqnarray} 
This confirms the validity of Eq. (114).

Having established that Eq.(109) and Eq.(114) are valid, then all the properties exhibited by $\lambda_I(E)$, in the one dimensional case (i.e. Sec. 4 on the quartic anharmonic oscillator) , apply for the multidimensional case, allowing the OPPQ-BM formalism to generate eigenenergy estimates and eigenenergy bounds.

\newpage

\section{References}
\noindent [1] Bender C M and Orszag S A 1999 {\it Advanced Mathematical Methods for Scientists and Engineers} (New York: Springer)\\
\noindent [2] Handy C R and  Bessis D 1985 {\em Phys. Rev. Lett.} {\bf 55} 931 \\
\noindent [3] Handy C R, Bessis D, Sigismondi G, and Morley T D 1988 {\em Phys. Rev. A} {\bf 37} 4557 \\
\noindent [4] Handy C R, Bessis D, Sigismondi G, and Morley T D 1988 {\em Phys. Rev. Lett.} {\bf 60} 253 \\
\noindent [5] Le Guillou J C and Zinn-Justin J 1983  {\em Ann. Phys. (N.Y.)} {\bf 147} 57 \\
\noindent [6] Kravchenko Y P, Liberman M A,  and Johansson B  1996 {\em Phys. Rev. A} {\bf 54} 287 \\
\noindent [7] Schimerczek C and Wunner G 2014 {\em Comp. Phys. Comm.} {\bf 185}  614 \\
\noindent [8] Handy C R 1987 {\em Phys. Rev. A} {\bf 36}, 4411 \\
\noindent [9] Handy C R 2001 {\em J. Phys. A} {\bf 34} L271\\
\noindent [10] Handy C R 2001 {\em J. Phys. A} {\bf 34} 5065\\
\noindent [11] Shohat J A and  Tamarkin J D, 1963 {\it The Problem of Moments} (American Mathematical Society, Providence, RI) \\
\noindent [12] Boyd S and Vandenberghe L 2004 {\it Convex Optimization} (New York: Cambridge University Press)\\
\noindent [13] Lasserre J-B  2009 {\it { Moments, Positive Polynomials and Their Applications}} (London: Imperial College Press )\\
\noindent [14] Chvatal V 1983 {\it Linear Programming} (Freeman, New York)\\
\noindent [15] Simon B 2008 {\em Proc. Symp. Pure Math. Amer. Math. Soc.} (Providence. R I ) 314.
\noindent [16] Handy C R and Vrinceanu D 2013 {\em J. Phys. A: Math. Theor. }  {\bf 46} 135202   \\
\noindent [17] Handy C R and Vrinceanu D 2013 {\em J. Phys. B: At. Mol. Opt. Phys.} {\bf 46} 115002 \\ 
\noindent [18] Handy C R 2020 {\em arXiv:2011.15011}.





\end{document}